# The propagation of light pollution in the atmosphere

P. Cinzano[*,1] and F. Falchi[1,2]

[1] *Istituto di Scienza e Tecnologia dell'Inquinamento Luminoso (ISTIL), Via Roma 13, I-36016 Thiene, Italy*

[2] *CieloBuio, Coordinamento per la protezione del cielo notturno, Osservatorio Astronomico "G.V.Schiaparelli" Via A. Del Sarto 3 - 21100 Varese, Italy*

**ABSTRACT**

Recent methods to map artificial night sky brightness and stellar visibility across large territories or their distribution over the entire sky at any site are based on the computation of the propagation of light pollution with Garstang models, a simplified solution of the radiative transfer problem in the atmosphere which allows a fast computation by reducing it to a ray-tracing approach. They are accurate for a clear atmosphere, when a two-scattering approximation is acceptable, which is the most common situation. We present here up-to-date Extended Garstang Models (EGM) which provide a more general numerical solution for the radiative transfer problem applied to the propagation of light pollution in the atmosphere. We also present the LPTRAN software package, an application of EGM to high-resolution DMSP-OLS satellite measurements of artificial light emissions and to GTOPO30 digital elevation data, which provides an up-to-date method to predict the artificial brightness distribution of the night sky at any site in the World at any visible wavelength for a broad range of atmospheric situations and the artificial radiation density in the atmosphere across the territory. EGM account for (i) multiple scattering, (ii) wavelength from 250 nm to infrared, (iii) Earth curvature and its screening effects, (iv) sites and sources elevation, (v) many kinds of atmosphere with the possibility of custom setup (e.g. including thermal inversion layers), (vi) mix of different boundary layer aerosols and tropospheric aerosols, with the possibility of custom setup, (vii) up to 5 aerosol layers in upper atmosphere including fresh and aged volcanic dust and meteoric dust, (viii) variations of the scattering phase function with elevation, (ix) continuum and line gas absorption from many species, ozone included, (x) up to 5 cloud layers, (xi) wavelength dependant bidirectional reflectance of the ground surface from NASA/MODIS satellites, main models or custom data (snow included), (xii) geographically variable upward light emission function given as a three-parameter function or a Legendre polynomial series. Atmospheric scattering properties or light pollution propagation functions from other sources can be applied too. A more general solution allows to also account for (xiii) mountain screening, (xiv) geographical gradients of atmospheric conditions, including localized clouds, (xv) geographic distribution of ground surfaces, but it suffers from too heavy computational requirements. Comparisons between predictions of classic Garstang models and EGM show close agreement for US62 standard clear atmosphere and typical upward emission function.

**Key words:** Atmospheric effects – site testing – scattering – light pollution – radiative transfer

## 1. INTRODUCTION

A worldwide growing interest on methods for monitoring and quantifying light pollution and its effects, in particular those on the night sky brightness and stellar visibility, is driven by the effort to preserve humanity's capability to perceive the universe beyond the Earth (e.g. Crawford 1989, Marin and Jafari 2007) and by the more general effort to preserve the purity of the nighttime environment and the health of the beings (animals, plants and humans) living in it (Longcore and Rich 2004, Navara and Nelson 2007). The environmental threat represented by the alteration of the natural quantity of light as a consequence of the introduction into the night environment of artificial light from nighttime outdoor lighting is faced by the enforcement of laws and rules at municipal, province, state or country levels. The main techniques to lower the effects of light pollution are well known, but improvements and additional prescriptions have been suggested (Falchi et al., 2011). The growing consciousness of the environmental consequences of light pollution leads to the integration of the monitoring of the artificial night sky brightness with specific quantification of the light pollution inside the atmosphere and on the ground surface. In recent years a series of works was published on the behaviour of light pollution in the atmosphere, notably those of Kocifaj (2007, 2008, 2011, 2012), Kocifaj, Aubé and Kohut (2010), Aubé (2007), Aubé and Kocifaj (2012).

---

[*] e-mail: cinzano@lightpollution.it

Falchi (1998) and Falchi & Cinzano (2000) used for the first time DMSP satellite imaging to compute maps of artificial and total sky brightness in large areas, obtaining maps of Italy's night sky using the Treanor (1971) propagation law. Cinzano et al. (2000), Cinzano, Falchi & Elvidge (2001a,b) and Cinzano & Elvidge (2003a,b,2004) presented methods to map across large territories the artificial night sky brightness, as well as the naked eye and telescopic limiting magnitude in a given direction of the sky and to compute the distribution of the night sky brightness and the limiting magnitude over the entire sky at any given site by evaluating the upward light emission from DMSP-OLS high resolution radiance calibrated data (Elvidge et al. 1999) and the elevation from GTOPO30 digital elevation map (Gesch, Verdin & Greenlee 1999). In these methods the computation of light pollution propagation in the atmosphere is based on the modelling technique introduced in 1986 by Roy Garstang and developed by him in the subsequent years (Garstang 1986, 1987, 1988, 1989a, b, 1991a, b, c, 2000, 2001). The Garstang modelling technique is fast, which is a main requirement in map computation and the comparison of predictions with observations has not evidenced so far any need to improve accuracy. However a comparison of the model detail with that of the models used in atmospheric physics for the computation of light propagation suggests making available to the scientific community an up-to-date extension of Garstang models. The price to pay for a more accurate behaviour is a slower code. Moreover, the greater detail introduced in the physical description is somehow counterbalanced by the approximations which must be introduced in the numerical computation due to the constraint on computational times. These limits will gradually disappear with time as faster computers become available.

In this paper we present up-to-date Extended Garstang Models (EGM) which provide a more general numerical solution for the radiative transfer problem applied to the propagation of light pollution in the atmosphere. We integrate the monitoring of the artificial night sky brightness with specific quantification of the light pollution of the atmosphere and at the ground surface. We also present the software package LPTRAN, an application of EGM to high-resolution DMSP-OLS satellite measurements of upward artificial light flux and to GTOPO30 digital elevation data, which provides an up-to-date method to predict the artificial brightness distribution of the night sky at any site in the World at any visible wavelength for a broad range of atmospheric situations. In sec. 2 we discuss the radiative transfer problem applied to the light pollution propagation in the atmosphere and present numerical solutions. In sec. 3 we describe the application to the computation of the observed artificial night sky brightness, the artificial radiation density in the atmosphere and the horizontal irradiance at ground level for a given atmospheric and surface situation. We also present the application to high-resolution satellite radiance measurements and digital elevation data. In sec. 4 we discuss the computation of detailed atmospheric and surface models. In sec. 5 we compare predictions of classical Garstang models and EGM, and present results of test applications. Conclusions are in sec.6.

## 2. RADIATIVE TRANSFER AND LIGHT POLLUTION PROPAGATION

### 2.1. Radiative transfer problem for light pollution propagation

Let's consider the 3-dimensional space where light pollution propagates. The coordinates are longitude, latitude and elevation, and the geometry is curved due to Earth curvature. The atmosphere has no plane parallel behaviour but can have gradients both with elevations and with geographic position. Incoming light rays are not parallel, differently from when sources are at infinity (like e.g. the Sun and the Moon) and light sources are under the atmosphere rather than over. This makes the situation very different from the daytime one. The intensity of light at any point depends on the geometrical position and the considered direction (given by three direction cosines or two position angles). The variation of the intensity along an infinitesimal path depends on how much radiation is scattered away or absorbed, how much radiation travelling in other directions is scattered into this direction and how much is produced by sources. The balance is regulated by the radiative transfer equation:

$$\frac{dI_\lambda}{dr}\mu = \beta I_\lambda - F_\lambda - S_\lambda, \qquad (1)$$

where $I$ is the energy per unit solid angle per unit time per unit surface, $\mu$ is the cosine of the angle between the light path for which I is evaluated and the direction $dr$ (the local vertical direction), $F$ is the energy per unit time per unit solid angle per unit volume scattered in this direction, $S$ is the source function which gives the light energy flux produced per unit solid angle per unit volume in this direction, $\beta$ gives the fraction of light scattered away or absorbed per unit length ($\beta$ is a scattering cross section per unit volume but it includes absorption, so it is more properly an exctinction cross section per unit volume or attenuation factor). If $dr$ is computed along the light path, then $\mu=1$. $I$, $F$ and $S$ depend on (x, y, z, $\theta$, $\phi$), the position and the zenith distance and azimuth of the direction of propagation of the light. We assume coherent scattering, so this equation can be solved individually for any wavelength, and the monochromatic solutions integrated or summed together.

The radiative transfer equation applied to the problem of the propagation of light pollution in the atmosphere is:

$$\frac{dI_\lambda(x,y,z,\theta,\phi)}{dr}\mu = I_\lambda(x,y,z,\theta,\phi)\sum_i \sigma_i n_i(x,y,z)$$
$$-\iint \sum_i \sigma_i n_i(x,y,z)\varpi_i \Phi_i(\theta',\phi') \quad (2)$$
$$\times \frac{1}{4\pi} I_\lambda(x,y,z,\theta',\phi')\sin\theta' d\theta' d\phi' - S_\lambda(x,y,z,\theta,\phi),$$

which, dividing for $\sum_i \sigma_i n_i(x,y,z)$, can also be written:

$$\frac{dI_\lambda(x,y,z,\theta,\phi)}{d\tau}\mu = I_\lambda(x,y,z,\theta,\phi)$$
$$-\iint \frac{\sum_i \sigma_i n_i(x,y,z)\varpi_i \Phi_i(\theta',\phi')}{\sum_i \sigma_i n_i(x,y,z)} \frac{1}{4\pi} \times \quad (3)$$
$$\times I_\lambda(x,y,z,\theta',\phi')\sin\theta' d\theta' d\phi' - \frac{S_\lambda(x,y,z,\theta,\phi)}{\sum_i \sigma_i n_i(x,y,z)},$$

where $\sigma_i$ are the extinction cross sections per particle of each species $i$, $n_i(x,y,z)$ are the particle densities of each species, $d\tau_\lambda = \sum_i \sigma_i n_i(x,y,z)dr$ is the optical depth due to all species. The single scattering albedo $\varpi$ measures the effectiveness of scattering relative to extinction and is defined as the ratio of the amount of flux scattered to that scattered and absorbed. The phase function $\Phi_i(\theta',\phi')$ gives the distribution of the photons scattered in different directions by each species. It is defined as the ratio of energy scattered per unit solid angle in the given direction to the average energy scattered per unit solid angle in all directions (Van de Hulst 1957). Then, if the scattering centers are randomly distributed and the scattering is coherent (i.e. if there is no wavelength change in the scattered radiation) the phase function has to obey the following normalization condition:

$$\frac{1}{4\pi}\int_0^{4\pi} \Phi_i(\theta',\phi')d\Omega = 1. \quad (4)$$

Some authors, like Garstang (1986, 1989, 1991a), define an angular scattering function $f = \frac{\Phi}{4\pi}$ which normalization condition is $\int_0^{4\pi} f d\Omega = 1$ so that its units are $sr^{-1}$. Other authors include the single scattering albedo in the phase function. The angles $\theta,\phi$ define the considered direction for which we want the intensity, whereas the angles $\theta',\phi'$ define the direction of the incoming light in respect to the previous direction. Given that the Earth curvature is not negligible on distance scales of some hundreds of kilometres at which light pollution propagates, the relationships between angles $\theta,\phi,\theta',\phi'$ are obtained from spherical and plane triangles.

The central part of eq. (2) gives the light flux scattered in the direction $(\theta,\phi)$ by a unit volume of atmosphere in $(x,y,z)$:

$$F_\lambda(x,y,z,\theta,\phi) = \iint \sum_i \sigma_i n_i(x,y,z)\varpi_i \Phi_i(\theta',\phi')$$
$$\times \frac{1}{4\pi} I_\lambda(x,y,z,\theta',\phi')\sin\theta' d\theta' d\phi'. \quad (5)$$

The integrand gives the fraction of incoming light which is scattered in the considered direction. The total phase function of the mix of species at $(x,y,z)$ is:

$$\Phi(x,y,z,\theta',\phi') = \frac{\sum_i \sigma_i n_i(x,y,z)\varpi_i \Phi_i(\theta',\phi')}{\sum_i \sigma_i n_i(x,y,z)\varpi_i}, \quad (6)$$

the total single scattering albedo is:

$$\varpi(x,y,z) = \frac{\sum_i \sigma_i n_i(x,y,z)\varpi_i}{\sum_i \sigma_i n_i(x,y,z)}, \quad (7)$$

and the total attenuation factor is (the reader should note that sometime $\beta = \frac{d\tau}{dz}$, but here we choose a more general form):

$$\beta(x,y,z) = \frac{d\tau}{dr} = \sum_i \sigma_i n_i(x,y,z). \quad (8)$$

They depend on the wavelength. Then the central part of eq. (2) can be rewritten:

$$F_\lambda(x,y,z,\theta,\phi) = \iint \beta(x,y,z)\varpi(x,y,z)\Phi(x,y,z,\theta',\phi')$$
$$\times \frac{1}{4\pi} I_\lambda(x,y,z,\theta',\phi')\sin\theta' d\theta' d\phi'. \quad (9)$$

### 2.2. Boundary conditions

We assume no celestial sources of light (although these could in principle be added as source terms, or the natural light could be computed separately and added), so at an arbitrary elevation the sky acts as a perfectly black surface. The source function $S_\lambda(x,y,z,\theta,\phi)$ is zero for $z > z_g$, where $z_g$ is the elevation over sea level of the Earth surface in $(x,y)$. In fact we assume that there are no sources of artificial light inside the atmosphere above ground level. Security lights of airplanes are not considered a main source of pollution and very polluting sources like advertising balloons or luminous satellites are a matter of other studies. No other sources exist, because natural light sources are not considered in this paper.

The source function $S_\lambda(x,y,z,\theta,\phi)$ is different from zero only on the Earth surface ($z=z_g$), where for zenith angles $\theta \leq \pi/2$ the intensity per unit surface is:

$$I_S(x,y,z_g,\theta,\phi) = e(x,y)\Im_{up,\lambda}(x,y,\theta,\phi) + \int_0^{2\pi}\int_0^{\frac{\pi}{2}} \cos\theta \cos\theta' \frac{1}{\pi}$$
$$\times I'_\lambda(x,y,z_g,\theta',\phi') R_\lambda(x,y,\theta,\phi,\theta',\phi') \sin\theta' d\theta' d\phi', \quad (10)$$

where $\Im_{up,\lambda}$ is the normalized upward light emission function of each source area, i.e. a normalized intensity so that its unit of measure is sterad$^{-1}$ (see also this paper in section 4.3 and Cinzano et al. (2000, sec. 4.3), $e(x,y)$ is the upward flux emitted per unit surface by each source area (Cinzano, Falchi & Elvidge (2001b, sec. 3.2), the integral gives the intensity of the light coming from the atmosphere and reflected from the ground surface, $I'_\lambda(x,y,z_g,\theta,\phi)$ is the energy per unit time per unit solid angle per unit surface incident on the surface. The Bidirectional Reflection Distribution Function (BRDF) $R_\lambda(x,y,\theta,\phi,\theta',\phi')$, defined in sec. 4.2, gives the relative intensity emitted by the ground surface in the direction of zenith angle $\theta$ and azimuth $\psi$, in function of the incident angles $\theta',\phi'$, which here are defined differently than in previous equations. No light is allowed to propagate below the Earth's surface.

## 2.3. Quantities of interest in light pollution studies

The basic needed information on the artificial light in the atmosphere is given by intensity per unit surface $I_S(x,y,z,\theta,\phi)$. The polluting effects of light pollution depend on the direction of the light. Only the intensity, which is a quantity depending on the direction, is the proper parameter to evaluate the effects of the artificial light emitted by a source, or, as in this case, coming from a volume of atmosphere centered in $(x,y,z)$. Quantities integrated on the sphere or on the upward and downward hemisphere miss this fundamental directional information, which cannot be discarded when the effects of light pollution are to be evaluated.

Quantification of light pollution means quantification of the alteration of the natural quantity of light in the night environment due to introduction of artificial light. When the atmosphere is considered not as the medium inside which light pollution propagates but as a part of the environment altered by light pollution, integrated quantities become useful as indicators of this alteration of the atmosphere because they summarize what would otherwise be too much detailed information. The primary indicators are (i) the *artificial night sky brightness (or radiance or luminance)* which indicates the integral of the artificial light scattered along the line of sight of an observer and has important effect on the perceived luminosity of the sky, on the star visibility, on the perception of the universe by mankind, on the darkness and the aspects of the environment, etc; (ii) the *sky irradiance or the sky illuminance on the Earth surface*, which has direct effects not only on the luminosity of the ground surface but also on the luminosity of the night environment as perceived by animals, plants and humans (of course where light pollution due to direct irradiance by nearby lighting installations is not overwhelming); (iii) the *radiation density* in the atmosphere, intended as the energy (or the light or the number of photons) per unit volume of atmosphere which is in course of transit, per unit time, in the neighborhood of the point $(x,y,z)$; (iv) the radiation density can be split in *upward and downward radiation densities*, which are useful to quantify approximately the light coming back toward the soil and the light going toward outer Space (we specified "approximately" because, due to the curvature of the Earth, not all the downward light goes on the ground); (v) another useful quantity is the *radiation density due to direct illumination* by the sources, i.e. the direct light travelling through a unit volume of atmosphere; (vi) finally of interest are the *upward and downward scattered flux densities* i.e. the flux density of the scattered radiation; the downward one, in particular, quantifies the "strength" of the unit volume of atmosphere at position $(x,y,z)$ as secondary source of light pollution when subjected to the considered light polluting action. It is opportune to stress again that the effects of the atmosphere as a secondary source of light pollution must be evaluated based on the intensity of light at each position whereas these integrated quantities are useful only as generic indicators of the alteration of the atmosphere itself.

In this paper we use as general terms radiance, irradiance and radiation density. In practical applications we will use (i) luminance, illuminance and luminous density for light in the CIE photopic and scotopic photometric bands, (ii) brightness and magnitude for light in astronomical photometrical bands, (iii) photon radiance, photon irradiance and photon density for radiation in other photometrical bands and (iv) spectral photon radiance, spectral photon irradiance and spectral photon density when considering spectral distribution of light pollution. However, following the traditional use in astronomy, we will call "night sky brightness" the flux of "anything" arriving by the night sky per unit surface per unit solid angle, independently by the considered constituents (e.g. energy, light, photons, etc.) and independently by the quantity effectively measured (e.g. radiance, luminance, astronomical brightness, photon radiance, etc.) or by its units (e.g. w m$^{-2}$ sr$^{-1}$, cd/m$^2$, mag/arcsec$^2$, ph s$^{-1}$ m$^{-2}$ sr$^{-1}$, etc.). $I_\lambda$ is given here for generality, but normally the correspondent quantity $I$, integrated over a given passband, is used.

The night sky brightness perceived by an observer on the Earth surface in $(x,y,z_g)$ looking in the direction of zenith angle and azimuth $(\theta,\phi)$ is simply $I_\lambda(x,y,z_g,\theta,\phi)$. The total horizontal irradiance $i_{\lambda,g}$ on the Earth surface is:

$$i_{\lambda,g}(x,y) = \int_0^{2\pi}\int_0^{\frac{\pi}{2}} I_\lambda(x,y,z_g,\theta,\phi) \cos\theta \sin\theta d\theta d\phi. \quad (11)$$

The downward radiation density $u_{\lambda,d}(x,y,z)$ is:

$$u_{\lambda,d}(x,y,z) = \frac{1}{c}\int_0^{2\pi}\int_{\frac{\pi}{2}}^{\pi} I_\lambda(x,y,z,\theta,\phi)\sin\theta\, d\theta\, d\phi. \quad (12)$$

In fact, if $i = \frac{dE}{dSdt}$ is the energy flux per unit surface per unit time at (x,y,z), the radiation density expressed as energy per unit volume is $\frac{dE}{dV} = \frac{dE}{dSdt} \times \frac{dt}{dr} = i/c$ where $c = \frac{dr}{dt}$ is the velocity of the light (see e.g. Chandrasekhar 1950 for a more detailed derivation). The radiation density can be expressed as energy density (energy per unit volume in J/m$^3$) or photon density (photons per unit volume in ph/m$^3$). The radiation density of light in CIE photopic and scotopic passbands can be expressed as luminous density (luminous energy per unit volume in Tb/m$^3$, where the talbot (Tb)= lumen × second is the unit of luminous energy). The upward radiation density is obtained integrating for $\theta$ between 0 and $\pi/2$.

The density $s_{\lambda,d}(x,y,z)$ of the flux scattered downward by a unit volume of atmosphere centered in $(x,y,z)$ is:

$$s_{\lambda,d}(x,y,z) = \frac{1}{c}\int_0^{2\pi}\int_{\frac{\pi}{2}}^{\pi} F_\lambda(x,y,z,\theta,\phi)\sin\theta\, d\theta\, d\phi. \quad (13)$$

The upward scattered flux is obtained integrating for $\theta$ between 0 and $\pi/2$. They are not a density of radiation (e.g. number of photons per unit volume) but a density of flux (e.g. number of photons per unit time per unit volume).

### 2.4. Previous solutions

A number of simplified solutions of the radiance transfer problem applied to light pollution were attempted by some researchers based on approximations or semi-empirical considerations. They were mainly interested to the artificial night sky brightness, so they searched for subsets of $I_\lambda(x,y,z_g,\theta,\phi)$ rather than for a general solution of the radiative transfer equation. First to try an approximate solution was Treanor (1973) who obtained a very simple solution which reasonably fitted observations of zenith night sky brightness at various distances from a city and was used for preparing the first map of artificial night sky brightness in Italy (Bertiau, Treanor & de Graeve 1973). Main assumptions were (i) homogeneous atmosphere, (ii) flat Earth, (iii) constant scattering coefficient, (iv) no ground reflection, (v) isotropic point source, (vi) single scattering. An empirically modified version of his formula was used by Berry (1976) and Pike & Berry (1978) to map artificial night sky brightness in Ontario, Canada. Garstang (1984) generalized Treanor's formula introducing a variable density of scattering particles along the vertical, which he assumed exponential, and a scattering coefficient depending on the scattering angle. Similar assumptions were taken later by Joseph, Kaufman & Mekler (1991) which added reflection by the Earth surface and accounted for different models of aerosols. A solution of radiative transfer equation using numerical integration was attempted by Yocke and Hogo (1986, eq.1) which assumed (i) homogeneous atmosphere, (ii) flat Earth,(iii) Lambertian ground reflection, (iv) phase function invariant in space, (v) isotropic point source. They accounted only for a single scattering but they tried a simplified treatment of multiple scattering based on considerations about energetic balance.

The main step in the solution of the radiance transfer problem applied to light pollution was taken by Garstang (1986, 1987, 1988, 1989a, b, 1991a, b, c, 2000, 2001). His work, based on a reasonably detailed physical model, produced a consistent follow-up due to the good accuracy and fast computation times which allow the application to the mapping of large territories, where the computation of a large number of individual contributions is requested. The main assumptions of Garstang Models are: (i) vertically inhomogeneous atmosphere with exponentially decreasing density and different scale height for aerosols and molecules; (ii) aerosol content set by the input atmospheric clarity parameter, related to the horizontal visibility or the stellar extinction; (iii) phase functions for aerosols and molecules invariant in space; (iv) curved Earth; (v) no ground reflection; (vi) point source with any upward intensity function (Garstang considered cities as circles of uniform brightness approximated with a grid of up to 52 points and used a 2-parameters upward emission function (Garstang 1989 eq.1); (vii) elevation of the site and the source over the ground plane; (viii) any wavelength for which the scattering sections are available (the author provided those for B and V photometric bands); (ix) double scattering via correction factor for both aerosols and molecules; (x) volcanic and desert dust layers (Garstang 1991a,b); (xi) cloud layer (Garstang 2001). We leave the readers to the cited papers for a more detailed understanding of Garstang models. Cinzano (2000) extended the model to triple scattering in order to compute light pollution around searchlight beams. Cinzano, Falchi & Elvidge (2001b) and Cinzano & Elvidge (2004) added account for mountain screening.

The Garstang modelling technique appears still accurate in common applications, given the uncertainties of available measurements. However there are good reasons to generalize it: (i) the atmospheric model, based on exponentially decreasing densities and an invariant phase function, appears too simple compared with the detailed models available in atmospheric physics; (ii) the ground reflection could be not negligible in some cases (in particular in case of snow cover), (iii) the approximate account for multiple scattering, particularly from Rayleigh scattering, become an important limit for unclean atmosphere or for particular sources (e.g. individual searchlights or projectors); (iv) the approximate account for continuous and line gas absorption limits accuracy when applied to spectra instead of wide photometric bands; (v) effects of multiple cloud layers and inversion layers are worth studying.

### 2.5. Numerical solution for the general 3D case

In order to extend the Garstang models to the more general radiation transfer problem for light pollution in the atmosphere, we searched for a numerical solution to the radiative transfer problem based on the following points:

i. the 3D space is divided in $i \times j \times k$ volumes of atmosphere, which volume depends on their elevation due to the curved geometry and on the subdivision. At the center of each volume a grid point is set, identified by a set of three indexes i,j,k.

ii. the Earth surface is divided in $i \times j$ land areas with the same size of the foot of the 3D columns. Each land area is assumed to have elevation $z_{gr}$ on the sea level. At the center of each land area there is a grid point, identified by a set of indexes i,j.

iii. the Earth surface is approximated to spherical in the considered area

iv. the array $F_{i,j,k,p,q}$ gives the quantity of light flux scattered per unit solid angle per unit volume of atmosphere in $(x_i, y_j, z_k)$ (above the atmosphere the scattering probability is zero). Indexes $p$, $q$ are a discretization, covering the entire sphere, of the angles identifying the direction of the emitted light ($4\pi$ steradians): the zenith distance $\theta(p)$ and the azimuth $\phi(q)$. The average phase function of a volume of atmosphere is usually continuous and quite smooth so it is sufficient to compute the values of F in a grid of values to be able to obtain values for others angles by interpolation. Then:

$$F_{i,j,k,p,q} = \\ \sum_{i',j',k'} \beta_{i,j,k} \varpi_{i,j,k} \Phi_{i,j,k}(\theta',\phi') \frac{1}{4\pi} i_\lambda(i,j,k,i',j',k') \\ + \sum_{i',j'} \beta_{i,j,k} \varpi_{i,j,k} \Phi_{i,j,k}(\theta',\phi') \frac{1}{4\pi} i_{\lambda,g}(i,j,k,i',j'),$$  (14)

where $i_\lambda(i,j,k,i',j',k')$ and $i_{\lambda,g}(i,j,k,i',j',k')$ are the irradiances in i,j,k from each other volume i',j',k' of atmosphere and each land area i', j' on the Earth surface. The irradiance $i$ replaces in the summation the integrand $I d\theta d\phi$ of eq. (9). The angles $\theta',\phi'$ depend on p,q and on the direction of the incoming light, i.e. on the geometrical relationship between the point i, j, k and i', j', k' or i', j'.

v. the array $T_{i,j,p,q}$ gives the quantity of light flux emitted per unit solid angle per unit surface (i.e. luminance) by the Earth surface in $(x_i, y_j, z_{gr})$. The BRDF is usually continuous and sufficiently smooth to allow computing the values of T in a grid of values and to obtain values for others angles by interpolation. Then:

$$T_{i,j,p,q} = \sum_{i',j',k'} \frac{1}{\pi} R(\theta,\phi,\theta',\phi') \cos\theta' i_{\lambda,g}(i,j,i',j',k') \\ + e_{i,j} \Im_{up,i,j},$$  (15)

where $i_{\lambda,g}(i,j,i',j',k')$ is the irradiance in i, j from each volume of atmosphere i', j', k', the $\cos\theta'$ term accounts for the horizontal irradiance, $e_{i,j}$ is the upward flux emitted by the land area (i, j), $I_{up,i,j}$ is the normalized upward emission function of the land area, $\theta,\phi,\theta',\phi'$ depend on p,q and on the direction of the incoming light, i.e. on the geometrical relationship between the point ($x(i), y(j), z_{gr}$) and ($x(i'), y(j'), z_{k'}$).

vi. In order to calculate the previous irradiances $i$, the following radiative transfer equation along each light path should be solved:

$$\frac{dI_\lambda}{dr} = \beta I_\lambda.$$  (16)

In fact inside the atmosphere it is S=0 and along the path it is F=0 because only the light coming from a specific volume or surface is considered. The solution of this simple equation is $I = I_0 \cdot e^{\int \beta dr}$ and can be obtained numerically, by summing $\Delta\tau = \beta \Delta r$ along the path from the grid point i', j', k' or the ground surface i', j' to the grid point i, j, k or the ground surface i, j:

$$i = F(\theta,\phi) \Delta V / r^2 e^{\sum_n \beta_n \Delta r_n}$$  (17)

$$i = T(\theta,\phi) \cos\theta \Delta S / r^2 e^{\sum_n \beta_n \Delta r_n},$$  (18)

where the first equation holds when the source is a volume of atmosphere and the second when the source is a land area on the Earth surface, $F(\theta,\phi)$ and $T(\theta,\phi)$ are obtained by interpolating $F_{i,j,k,p,q}$ and $T_{i,j,k,p,q}$ in the direction ($\theta',\phi'$) of the light path, $\Delta V$ and $\Delta S$ are the volume and the area of the secondary source. Note that in a curved geometry the size of the volumes of atmosphere over the same area increases with their elevation. The attenuation factor $\beta_n$ is computed at the center of each integration interval $\Delta r_n$. Eqs. (17) and (18) provide a simple way to account for the mountain screening. It is sufficient to set $I_0 = 0$ when along the light path $z'' \leq z_{gr}$.

The solution can be computed by applying the following computational sequence, where $F^r$, $T^r$ are the r-th estimate in the sequence: (i) computation of $F^0$ array for each volume of atmosphere based on $T^0$ array, where only upward emission is accounted at beginning; (ii) computation of $F^1$ array of each volume based on $F^0$ arrays of the other volumes and the $T^0$ array (this corresponds to taking into account two scatterings); (iii) computation of $T^1$ array of each land surface based on $F^0$ array of the volumes of atmosphere (again two scatterings); (iv) computation of $F^{r+1}$ array of each volume based on $T^r$ array of each land surface and $F^r$ array of the other volumes (this corresponds to taking into account three scatterings); (v) computation of $T^{r+1}$ array of each land surface based on $F^r$ array of the volumes of atmosphere; (vi) continue iterations until the wanted number of multiple scatterings is accounted for. The night sky brightness obtained from the r-th estimate accounts for r+1 scatterings. The intensity $I_\lambda$, which is the true solution of the radiative transfer equation, is not saved in an array. In fact the intensity at any grid point (i, j, k) is strongly peaked in the direction opposite to sources, differently from the scattered light which is quite smooth, so that it would be necessary to have a very dense array in p,q to accurately store it. Moreover the light intensity coming from near volumes, considered in this scheme as point sources, would produce errors in the intensity distribution, whereas the error produced in the scattered intensity, which is smooth, is much smaller. Given that only $F_{i,j,k,p,q}$ is required to compute the artificial night sky brightness, to limit computational requirements and computational time, we can store only it, together with the T array and the integrated upward and downward light densities at any grid point. The array of the extinction between pairs of volumes and surfaces could be saved and used again, however its size is the square of the grid points $(i \times j \times k+1)^2$, very huge for non-dedicated computers, so it is less expensive computing it each time.

In summary, the input data are: (i) the size of the considered area, maximum height and grid set up; (ii) the 3D distribution of the total attenuation factor; (iii) the 3D distribution of the total phase function; (iv) the 3D distribution of the single scattering albedo; (v) the 2D digital elevation map; (vi) the 2D distribution of sources and their upward light emission functions; (vii) the 2D distribution of surface bidirectional reflectance.

2.5.1. Indicators of light pollution

We do not store the intensity per unit surface $I_\lambda(x,y,z,\theta,\phi)$ in an array, so the night sky brightness $I_\lambda(x,y,z_{gr},\vartheta,\varphi)$ perceived by an observer on the Earth surface looking in the direction $(\vartheta,\varphi)$ should be obtained integrating the light emission of each volume of atmosphere along the line of sight, accounting for radiative transfer along the light path. This requires solving the equation:

$$\frac{dI_\lambda}{dr} = \beta I_\lambda - F_\lambda, \qquad (19)$$

where now F is a known function of the position and the direction and accounts for multiple scattering and reflection. This linear differential equation of first order has the solution:

$$I_\lambda(x,y,z_{gr},\vartheta,\varphi) = \int_0^\infty e^{-\int_0^\ell \beta dr} F_\lambda(x',y',z',\theta',\phi')d\ell, \qquad (20)$$

where $\beta$ and $F_\lambda(x',y',z',\theta',\phi')$ are functions of the position $\ell$ along the path and the direction of observation and the relations between the angles $\vartheta,\varphi,\theta',\phi'$ are given by spherical and plane trigonometry. In practice, we iterate the following equation from the last point of the line-of-sight still inside the grid to the observer:

$$I' = e^{-\beta_{i',j',k'}\Delta\ell}I + F_{i',j',k'}(\theta',\phi')\Delta\ell / r^2 e^{-\beta_{i',j',k'}\Delta\ell/2}, \qquad (21)$$

where F and $\beta$ are taken at the center of each $\Delta\ell$ interval. $F$ is approximated to the nearest grid point and it is interpolated on the direction of the light path. The irradiance on the Earth surface, the radiation densities and the scattered flux densities are obtained by numerical integration of eqs. (11), (12) and (13).

The limitation of the approach discussed above is that to be accurate it requires a dense grid (small volumes). This implies extremely huge arrays and very long computational times. So at the moment it appears reserved for dedicated computational facilities with adequate power capabilities, if any. Hence we searched for a further solution, computable in reasonable time with a common fast workstation of the last generation. This solution is presented in next section.

*2.6. Numerical solution for the case of axial symmetry*

To simplify the computation we consider only one source. The contributions of more sources can be added later because light pollution and night sky brightness are additive. We assume that the source emission has axial symmetry. This is a reasonable assumption, at least in densely populated areas, where the night sky brightness is produced by the sum of many cities and towns because the random distribution of asymmetries and differences in their upward emissions results in a very smoothed and symmetric average upward emission function. This assumption may fall off, especially in proximity of a single main source and particularly so if this source has, for example, roads in north-south and east-west directions, as in the case of several US cities. This axial asymmetry may explain the degree of polarization of the sky glow (Kyba et al. 2011). We assume that the atmosphere is always the same in the considered area (e.g. that no localized clouds are present but only extended clouds layers) or, at least, that it changes radially with the distance from the source (this is the case e.g. when there are clouds just over the source). We also assume that the reflectance properties of the ground surface are constant on average in the considered area, or that they change radially. With these assumptions the problem takes axial symmetry. This means that the spatial distribution of the light scattered by unit volume of atmosphere at distance x(i) and elevation z(j) is the same on any vertical plane passing through the source. Hence we can compute only $F_{i,j}(\theta,\phi)$ in place of $F_{i,j,k}(\theta,\phi)$, saving a lot of computations. As an example, in order to fill up to a distance of 350 km from the source with 1 km step, a 3D square grid centered on the source should have $700^2$ columns. If there are $n$ volumes per column, the array $F$ should store the radiance of $700^2 \times n$ volumes for $p \times q$ directions. The irradiance produced on each of these volumes by the other volumes should be computed $700^4 \times n^2$ times. Nevertheless, a semicilindrical grid with axial symmetry with a step of 5 degrees in azimuth requires to store the radiance in the array $F$ only for $350 \times n$ volumes and to compute the irradiance only $350^2 \times n^2 \times 37$ times. The computational times are roughly reduced by a factor $5 \cdot 10^4$ (e.g. a computation lasting 137 years would be reduced to a day). Moreover there is specular symmetry between the left and the right of the plane. Then, computing $F_{i,j,k}(\theta,\phi)$ for $0 \le \phi \le 180$, we also obtain $F_{i,j,k}(\theta,180-\phi)$. This could approximately halve the computational times, depending on how the code is written.

Then we searched for a numerical solution of eq. 2 based on the following points:

i. the problem has axial symmetry

ii. the Earth surface is assumed to be spherical

iii. in the 3D space are identified $i \times j \times k$ volumes of atmosphere in a cylindrical reference system. At the center of each volume there is a grid point. The index $i$ identifies the distance from the source on the horizontal plane, $j$ the azimuth angle of the grid line and $k$ the height.

iv. the Earth surface is divided in $i \times j$ land areas in a circular reference system, with the same size of the foot of the 3D columns. As before, each land area is assumed to have elevation $z_{gr}$ on the sea level. At the center of each land area there is a grid point, identified by indices i,j.

v. the array $F_{i,k,p,q}$ gives the quantity of light scattered per unit solid angle per unit volume of atmosphere in $(x_i, z_k)$. Due to axial symmetry the volumes are identical for rotation of a vertical plane passing through the source around the normal to the source. Indices $p$, $q$ are a discretization, covering the entire sphere, of the angles identifying the direction of the emitted light ($4\pi$ steradians): the two angles $\theta$ and $\phi$ are the zenith distance measured from the upward normal to the surface and the azimuth measured on a perpendicular plane from direction opposite to the source. As before, the average phase function is sufficiently smooth to allow us to obtain values for other angles by interpolation. Then:

$$F_{i,k,p,q} = \sum_{i',j',k'} \beta_k \varpi_k \Phi_k(\theta',\phi') i_\lambda(i,k,i',j',k') \quad (22)$$

or

$$F_{i,k,p,q} = \sum_{i',j',k'} \beta_k \varpi_k \Phi_k(\theta',\phi') i_{\lambda,g}(i,k,i',j'), \quad (23)$$

where $i_\lambda(i,k,i',j',k')$ is the irradiance at i, k from each other volume i', j', k', $i_{\lambda,g}(i,k,i',j')$ is the irradiance at i, k from the ground surface i', j', and $\Delta\theta, \Delta\phi$ depend on p,q and on the direction of the incoming light, i.e. on the geometrical relationship between the point i, k and i', j', k' or i', j'.

vi. the array $T_{i,p,q}$ gives the quantity of light emitted per unit solid angle per unit surface by the land area in $(x_i, z_{gr})$. Again, the average BRDF is usually continuous and sufficiently smooth to allow computing the values of T in a grid of values p,q and to obtain values for others angles by interpolation. Then:

$$T_{i,p,q} = \sum_{i',j',k'} \frac{1}{\pi} R(\theta,\Delta\phi,\theta') \cos\theta' i_{\lambda,g}(i,i',j',k') + e I_{up}(\theta), \quad (24)$$

where $i_{\lambda,g}(i,i',j',k')$ is the irradiance at the land area $i$ from each volume of atmosphere i', j', k' and $\theta, \theta', \Delta\phi$ depend on p,q and on the direction of the incoming light, i.e. on the geometrical relationship between the point $(x(i), z_{gr})$ and $(x(i'), y(j'), z_{k'})$, easy to obtain with some plane and spherical trigonometry. Here we have only one source at center, so only one upward flux $e$ and upward emission function $I_{up}(\theta)$, depending in turn only on the zenith angle. In the computation of the libraries of sec. (1.3) we assumed unit upward flux (e=1).

vii. As discussed in sec. 2.5, the irradiance $i$ can be obtained numerically by summing $\Delta\tau = \beta\Delta r$ along the path from the grid point i', j', k' or from the ground surface i', j' to the grid point i, k or the ground surface i:

$$i = F(\theta,\phi)\Delta V / r^2 e^{\sum_n \beta_n \Delta r_n} \quad (25)$$

$$i = T(\theta,\phi)\cos\theta \Delta S / r^2 e^{\sum_n \beta_n \Delta r_n}, \quad (26)$$

where the first equation holds when the source is a volume of atmosphere and the second when the source is a land area on the Earth surface, $F(\theta,\phi)$ and $T(\theta,\phi)$ are obtained by interpolating $F_{i,k,p,q}$ and $T_{i,p,q}$ in the direction $(\theta',\phi')$ of the light path, $\Delta V$ and $\Delta S$ are the volume and the area of the secondary source. Due to the curved geometry the size of the volumes of atmosphere increases with their elevation and due to the cylindrical coordinate system it also increases with distance from the center. The attenuation factor $\beta_n$ is computed at the center of each integration interval $\Delta r_n$. The screening due to Earth curvature is accounted for by setting $I_0 = 0$ when $z'' \le 0$ along the light path and $T_{i,p,q} = 0$ when $\theta_p \ge 90$ degrees.

The solution can be computed by applying the same iterative computational sequence described in sec. 2.5 for the 3D case. As discussed in sec. 2.5.1, the night sky brightness perceived by an observer on the Earth surface is obtained with eq. (21) from the array F whereas the irradiance on the Earth surface, the radiation densities and the scattered flux densities are obtained by numerical integration of eqs. (11), (12) and (13).

In summary, the input data are: (i) the size of the considered area, maximum height and grid data; (ii) the vertical distribution of the total attenuation factor; (iii) the vertical distribution of the total phase function; (iv) the vertical distribution of the single scattering albedo; (v) the source elevation; (vi) the upward light emission function; (vii) the average bidirectional reflectance distribution function of the ground surface.

## 3. APPLICATION

The software package LPTRAN (which stands for Light Pollution radiative TRANsfer), written in Fortran-77, applies the method described above for the case of axial symmetry. It is composed by a number of programs: the main program LPTRAN (the same name of the package) computes the radiative transfer and light pollution propagation based on an input atmospheric and surface model for the given wavelength, prepared as described in sec. 4.1, LPDART evaluates light pollution and night sky brightness on the grid and prepares a library of light pollution propagation data. The program LPSKYMAP_LPTRAN computes night sky brightness in a site based on DMSP-OLS radiance data, a Digital Elevation Map and the LPTRAN library. The LPSKYMAP package (Cinzano & Elvidge 2004) allows completion of the study. The programs LPSKYALT, LPSKYDENS and LPSKYFRZH compute across a territory the artificial night sky brightness at any azimuth and elevation, the radiation and scattered flux densities in the atmosphere and their fractional contribution to the zenith night sky brightness at sea level.

### 3.1. Radiative transfer and light pollution propagation

The computation of the radiative transfer and light pollution propagation is carried out by the program LPTRAN. We divided a column of atmosphere in layers and set up the grid of LPTRAN such that the elevation of each grid point corresponds to the centre of these layers. For each wavelength of interest and for each layer $k$ we need: (i) the elevation of its borders; (ii) the total vertical optical depth of the layer from which the average optical depth per unit length (average attenuation factor) is obtained; (iii) the average single scattering albedo, which is the probability that an extinction event scatters rather than absorbs a photon; (iv) the average total phase function $\Phi_k$ given by the first 20 coefficients of its Legendre polynomial expansion (according with DISORT conventions the coefficients are divided by 2n+1):

$$\Phi_k(\theta) = \sum_{n=0}^{20}(2n+1)p_n P_n(\cos\theta), \quad (27)$$

where $P_n(\cos\theta)$ are the Legendre polynomials of degree n and $p_n$ are the coefficients which can be obtained by integrating, possibly numerically, the equation:

$$p_n = \frac{1}{2}\int_0^\pi \Phi_k(\theta)P_n(\cos\theta)\sin\theta d\theta. \quad (28)$$

We also need the elevation of the source, the upward light emission function of the source at that wavelength and the bidirectional reflectance of the ground surface. They are obtained as explained later in sec. 2. The program LPTRAN calculates the arrays of the scattered light due to secondary scattering, the upward and downward flux density per unit volume at each grid point on the vertical plane passing for the source. It does not account for the refraction of light by the atmosphere which could increase the brightness near the horizon in sites polluted by very distant cities. We plan to account for this in further papers.

LPTRAN allows resumption of the computation after an unexpected computer stop, which is useful in long calculations. Typical calculation times on a workstation powered by a CPU AMD 64 3700+ for a grid of 150 points in distance, 33 volumes in height, 10step in azimuth, computing F every 10in zenith distance and azimuth, are of the order of 4.4 days each scattering. A few scatterings are usually sufficient for the purposes. Results for 30 different site elevations can be obtained in a few hours from LPTRAN results using LPDART, described in the next section. Hence a library of light pollution propagation functions for one atmosphere, one wavelength, one upward function, one ground reflectance, and one source elevation can be made in less than a month by a single workstation.

### 3.2. Light pollution evaluation

The program LPDART (which stands for Light Pollution Distribution from Atmospheric Radiative Transfer) calculates at the chosen wavelength, based on LPTRAN results, (i) the total artificial radiation density in the atmosphere as a function of elevation and distance from the source; (ii) the observed artificial night sky brightness in function of altitude, direction of observation and distance from the source, by integrating the light coming from each volume of atmosphere along the line of sight with position angles ($\vartheta$, $\varphi$) of an observer at $(x, y, z)$. The luminance in photopic or scotopic bands or the brightness in astronomical photometrical bands can be obtained with good approximation from the radiance at the effective wavelength or, with better accuracy, computing the radiance in the range from 1% to 1% of the band and integrating along the wavelength with the passband as weight.

Main LPDART results for each wavelength, atmospheric model, distance and elevation on the grid are:

(i) the array of the scattered light per unit solid angle per unit volume in a direction of given azimuth and zenith distance by each atmospheric volume;

(ii) the upward and downward scattered flux density by each atmospheric volume and the upward scattered flux

surface density at the ground surface;
(iii) the upward and downward radiation density at each atmospheric volume and the horizontal irradiance on the ground surface, due to the secondary sources (the atmosphere and the surface);
(iv) the radiation density in each atmospheric volume, due to direct illumination by the polluting source;
(v) the fractional contribution of each unit volume of atmosphere to the night sky brightness at zenith at sea level;
(vi) the light pollution propagation function $f$, i.e. the artificial night sky brightness $b$ produced by an individual source of unitary upward flux $e$ on a grid of zenith distances and azimuth angles for an observer at each distance from the source along the grid and for each observer elevation between the ground level and a maximum altitude over sea level (to limit computational times we usually take a maximum elevation of 3000 m with a 100 m step), after eq. (20) and sec. 2.3:

$$f = \frac{b}{e} = \frac{\int_{\lambda_0}^{\lambda_1} S_\lambda I_\lambda(x,y,z,\theta,\phi) d\lambda}{e}$$

where $S_\lambda$ is the weight function of the considered band (e.g. Johnson 1955).

### 3.3. Mapping the artificial night sky brightness in a site

Following Cinzano & Elvidge (2004), we prepared a modified version of their program LPSKYMAP called LPSKYMAP_LPTRAN. It computes the brightness distribution over the entire night sky hemisphere at any site in the World based on (i) high-resolution DMSP-OLS satellite measurements of upward artificial light emission (Elvidge et al. 1999), (ii) GTOPO30 digital elevation data (Gesch, Verdin & Greenlee 1999) and (iii) libraries of propagation of light pollution computed by LPTRAN/LPDART for the chosen atmospheric and surface model. Its results can be analyzed with the other programs of LPSKYMAP package (Cinzano & Elvidge 2004) to obtain the total night sky brightness and the star visibility (limiting magnitude), to compare results with available measurements, make polar or linear plots etc. The main difference between LPSKYMAP_LPTRAN and LPSKYMAP is that rather than computing the night sky brightness in a given direction of the sky, at a given distance from a source, for given elevations of the site and the source (called *light pollution propagation function* in Cinzano et al. 2000) based on classic Garstang models, it interpolates the value from a library computed by LPTRAN/LPDART for the chosen atmospherical/surface model.

A library of precalculated propagation functions for the chosen atmospheric conditions is required because computation times prevent making individual computations for each source area when mapping urbanized territories. As an example, in order to map the zenith night sky brightness in Europe with the grid sizes adopted by Cinzano, Falchi, Elvidge (2001a), assuming a filling factor of lighted areas of 1/4 (one land area is source of light every 4), are required about 4800x4800x300x300x0.25 individual computations. Only the use of precalculated libraries permits it. Then we subdivided the upward light emission function in some basic functions (discussed later in sec. 2.3), we discretized the source elevation and for each of them we ran LPTRAN and LPDART obtaining one library of functions. A library $f$ should be computed for each atmospheric and surface model, and for each wavelength. The artificial night sky brightness computed by LPSKYMAP_LPTRAN is (after Cinzano & Elvidge 2004):

$$b_{i,j} = \sum_h \sum_l e_{h,l} f_{n,m,r,w}(\theta_i, \phi_j, d_{h,l}, h_0, s_{h,l}), \qquad (29)$$

where $e_{h,l}$ is the upward flux emitted by the land area (h, l) and $f$ is the light pollution propagation function from the library. The indices i, j define the direction of observation. The indices n,m,r,w identify respectively the atmospheric model, the component of the upward emission function, the ground surface, the wavelength, $s$ is the source elevation from GTOPO30 digital elevation data, $\theta, \phi$ are respectively the zenith distance and azimuth (computed from the source direction) of the line-of-sight, $d$ is the distance from the source, depending on (h, k) and $h_0$ is the site altitude from the elevation data. LPSKYMAP_LPTRAN interpolates over $\theta, \phi$ if the library grid is coarser than its grid, does not interpolate over $d$ because the grid is assumed to be the same and takes the nearest grid point for $h$ because the step is small. The summations are extended to all the land areas around the site inside a distance for which their contributions are non-negligible or the library is available. Each run of LPTRAN and LPDART produces one file $b_{n,m,h,w}$ for each source elevation. Due to the computation times discussed in previous sections, the set up of a library for more wavelengths and atmospheric models is not a quick operation. The program LPSKYMAP_LPTRAN can also use libraries provided by any other programs or models, including classic Garstang models. Libraries should be given for each source elevation s as a binary array $b_s(naz, nh, nx, nsit)$, where naz is the number of azimuth angles, nh is the number of zenith distances, nx the number of distances from the source, nsit the number of site elevations. The grid should be adapted.

In summary, the LPSKYMAP_LPTRAN input data for night sky brightness computation at any site for a given atmospheric and surface model are:
(i) the light pollution propagation library;
(ii) an input file with the geographical position and elevation of the site, the names of input DEM and lights frames and the position of their upper left corner;
(iii) subimages with Digital Elevation map and DMSP-OLS radiance data cropped from the original large scale frames with the program MAKEFRAC and MAKESHIFT;
(iv) the calibration constant, based on pre-flight calibration at 1996-1997, on the Cinzano, Falchi & Elvidge (2001b) calibration at 1998-1999 from Earth-based measurements, or on observations taken at the same site analyzed with LPSKYCAL and LPSKYCOMPARE. The LPSKYMAP_LPTRAN program and the LPSKYMAP package produce a series of calibrated maps of the total night sky brightness, interpolated or not, and the limiting magnitude. They also add the horizon line. Maps in polar coordinates are obtained with the program LPSKYPOLAR. Comparison with observations is made with the program LPSKYCOMPARE. The programs MAKEFRAC, MAKESHIFT, LPSKYPOLAR, LPSKYCAL and LPSKYCOMPARE belong to the LPSKYMAP package (Cinzano & Elvidge 2004). Maps are analyzed with FTOOLS developed by HEASARC at the NASA/GSFC. Statistical analysis is made with the software MATHEMATICA of Wolfram Research.

The screening due to Earth curvature is accounted for but not the screening due to mountain or terrain elevation. The axial symmetry allows only for accounting for screening by Earth curvature because the screening by mountain or terrain requires information which is not axisymmetric but bidimensional: the elevation of each area of territory. While waiting for more computer power allowing accurate 3D solutions with mountain screening, possible ways to approximately estimate the screening effects are: (i) evaluation of their magnitude order at a site comparing the night sky brightness with and without mountain screening, as obtained with Classical Garstang Models, i.e. with the original LPSKYMAP program; (ii) considering that main effects of screening are on the direct illumination by the source, it could be possible updating LPSKYMAP_LPTRAN in order to subtract from the resulting sky brightness obtained neglecting mountain screening the fraction due to direct light screened by mountain and computed like in classic Garstang models. In both cases these corrections should improve somewhat the accuracy of the results.

The LPSKYMAP_LPTRAN program is faster than LPSKYMAP, because it does not need to compute the light pollution propagation and mountain screening. However, due to the computational time required by LPTRAN for computing each library, we usually compute individual maps rather than the hypermaps described by Cinzano & Elvidge (2004) which account for a range of aerosol contents and wavelengths. The expected rapid increase of low cost computational power in future years will allow it.

### 3.4. Mapping the artificial night sky brightness across a territory

The implementation of the LPTRAN libraries inside the methods for mapping the night sky brightness in a given direction across large territories (Cinzano et al. 2000, Cinzano, Falchi & Elvidge 2001a,b and following) is similar to the previous one, when no mountain screening is accounted for. This is the most common case because, even when faster Garstang models are used, the mountain screening can be accounted for only when small areas are mapped (few tens of kilometers of side) due to computational times. A map of the artificial night sky brightness $b_{i,j}$ in the chosen direction $(\theta_0, \phi_0)$ of the sky is computed by the program LPSKYALT from (after Cinzano, Falchi & Elvidge 2001b):

$$b_{i,j} = \sum_h \sum_l e_{h,l} f_{n,m,r,w}(\theta_0, \phi_0, d_{i-h, j-l}, h_{i,j}, s_{h,l}), \quad (30)$$

for each pair (i, j) and (h, l), which define the positions of the observing site and the polluting area on the array. The observing site is assumed at the centre of each land area (i, j) in which the territory has been divided. Here $e_{h,k}$ is the upward flux of the land area (h,k) obtained from satellite measurements as in Cinzano, Falchi & Elvidge (2001b) following the same calibration method, $f$ is the light pollution propagation library for the chosen atmosphere n, upward emission function m, ground surface r, wavelength w, source elevation $s$ and site elevation $h_{i,j}$ from digital elevation data. The upward emission function can vary with the land area giving for each of them the parameters of the 3-parameter function or the Legendre expansion.

### 3.5. Mapping the radiation density in the atmosphere

Following the computational scheme used above to map the night sky brightness across a territory, the program LPSKYDENS calculates the artificial upward/downward radiation density and the upward/downward scattered flux density on a 3D grid filling the atmosphere over a territory. The grid is given by the centers (i, j) of the land areas in which the territory is divided and by the k centers of the volumes in which each column of atmosphere is divided. The libraries $u_{n,m,r,w}(d,h,s)$ of radiation densities and scattered flux densities in the territory around a source are computed by LPTRAN and LPDART. The program LPSKYDENS computes for each grid point (i, j, k):

$$u_{i,j,k} = \sum_h \sum_l e_{h,l} u_{n,m,r,w}(d_{i-h,j-l}, h_k, s_{h,l}), \quad (31)$$

where the upward flux $e_{h,l}$ is obtained from DMSP-OLS radiance data as in Cinzano, Falchi, Elvidge (2001) following the same calibration method, $d$ is computed from the relative geographical positions and the other quantities have been defined in previous sections. The program interpolates linearly the library elements. The 3D arrays are saved as AGL files to be analyzed with the Compaq Array Viewer. Similarly, the program LPSKYFRZH computes the array of the fractional contribution of the light scattered by each volume to the artificial night sky brightness at zenith at sea level, based on the library $frzh_{n,m,r,w}(d,h,s)$ calculated by LPDART. It accounts for the extinction along the path from the volume to the ground and it refers only to the light scattered along the nadir direction.

### 4. INPUT DATA

### 4.1. Atmospheric model

The atmospheric scattering data required by LPTRAN, i.e. the vertical profiles of the total vertical optical depth, the average total single scattering albedo and the Legendre expansion coefficients of the average total phase function, can be obtained from any atmospheric scattering modelling software. We obtained them with the subroutines of the package SBDART (Santa Barbara DISORT Atmospheric Radiative Transfer), a Fortran-77 computer code for the analysis of a variety of radiative transfer problems encountered in satellite remote sensing and atmospheric radiation budget studies in a vertically inhomogeneous atmosphere (Ricchiazzi et al. 1998). SBDART is based on a collection of well-tested and reliable physical models which have been developed by the atmospheric science community over the past few decades (Ricchiazzi et al. 1998). Thanks to their modularity its subroutines provide a good starting point for researchers interested in developing their own radiation-transfer codes. Here we used mainly SBDART version 1.21 (1998) except for sea reflectance which was taken from release 2.4 (2002).

We wrote a modified version of the main program SBDART which sets up and saves in a file the atmospheric vertical profiles at each wavelength in a chosen range and with a chosen step. SBDART computes the average between molecules, aerosols, including molecular and gas absorption. The routine DISORT (DIScreet Ordinate Radiative Transfer) which numerically solves the equations of plane-parallel radiative transfer, is not used because it is not of interest in this case.

Here we shortly summarize the main characteristics of the model atmosphere made by SBDART following Ricchiazzi et al. (1998). We refer the readers to these authors for a more detailed presentation and to Ricchiazzi (2002) for a detailed documentation about input parameters.

### 4.1.1. Standard Atmospheric Models

SBDART is configured to allow up to 40 atmospheric layers. It adopts six standard atmospheric profiles from the 5s atmospheric radiation code which are intended to model the following typical climatic conditions: tropical, midlatitude summer, midlatitude winter, subarctic summer, subarctic winter and US62 standard atmosphere. These model atmospheres (McClatchey et al. 1972) have been widely used in the atmospheric research community and provide standard vertical profiles of pressure, temperature, water vapor and ozone density. The concentration of trace gases such as $CO_2$ or $CH_4$ are assumed to make up a fixed fraction of the total particle density. In addition, the user can specify its own atmosphere by listing, for each elevation, the pressure, the temperature, the water vapor density, the ozone density and by setting many optional parameters, like the volume mixing ratio of a number of species (default values in parenthesis): $O_2$ ($2.09 \cdot 10^5$ ppm), $CO_2$ ($3.3 \cdot 10^2$ ppm), $CH_4$ (1.74 ppm), $N_2O$ (0.32 ppm), CO (0.15 ppm), $NH_3$ ($5.0 \cdot 10^{-4}$ ppm), $SO_2$ ($3.0 \cdot 10^{-4}$ ppm), NO ($3.0 \cdot 10^{-4}$ ppm), $HNO_3$ ($5.0 \cdot 10^{-5}$ ppm), $NO_2$ ($2.3 \cdot 10^{-5}$ ppm). This makes possible the use of the true local atmospheric conditions at the same time of satellite measurements or the typical local atmospheric conditions in a clear night, when they will be available on a global scale.

### 4.1.2. Aerosol Models

The aerosol models included in SBDART are derived from those provided in the 5S CODE (Tanre et al. 1990) and LOWTRAN7 code (Shettle & Fenn 1975). SBDART accounts for several common boundary layer and upper atmosphere aerosol types. In the boundary layer, either rural, urban, oceanic and tropospheric aerosols can be selected, which differ from one another in the way their scattering efficiency, single scattering albedo and asymmetry factors vary with wavelength. The spectral dependence of the aerosol scattering parameters can be also specified by the user. The total vertical optical depth of boundary layer aerosols is derived from user specified horizontal meteorologic visibility at 550 nm and an internal vertical distribution model. Visibility is defined in SBDART as the horizontal distance in km at which a beam of light at 550 nm is attenuated by a factor of 0.02:

$$V = \frac{-ln(.02)}{n(0)\sigma}, \qquad (32)$$

where $\sigma$ is the aerosol absorption+scattering cross-section at 550 nm (AMS, 1959) and $n(0)$ is the number density of particles. It corresponds to Garstang (1989) eq.38.

In the upper atmosphere up to 5 aerosol layers can be specified based on their altitude above the surface, with radiative characteristics that model fresh and aged volcanic dust, meteoric dust and stratospheric background aerosols.

### 4.1.3. Molecular Absorption Model

Gas Absorption Model of SBDART relies on low-resolution band models developed for the LOWTRAN-7 atmospheric transmission code (Pierluissi & Marogoudakis 1986). These models provide the clear sky atmospheric transmission from 0 to 50000 $cm^{-1}$ and include the effects of all radiatively active molecular species found in the Earth's atmosphere. The models were derived from detailed line-by-line calculations which were degraded to 20 $cm^{-1}$ resolution for use in LOWTRAN. This translates to a wavelength resolution of about 5 nm in the visible and about 200 nm in the thermal infrared.

### 4.1.4. Cloud Model

As pointed out by Kyba et al. (2011) and by Lolkema et al. (2011), cloud coverage may dramatically increase the sky luminance, especially near or inside cities. This effect is expected to increase the ecological impact of light pollution. To allow simulation of this effect, up to 5 cloud layers can be defined. For modelling clouds SBDART uses the Henyey-Greenstein parameterization of the scattering phase function. This approximation depends only on the asymmetry factor, which indicates the strength of forward scattering, and has been shown to provide good accuracy when applied to radiative flux calculations (van de Hulst 1968; Hansen 1969). SBDART computes the scattering efficiency, the single scattering albedo and the asymmetry factor within a cloudy atmosphere using a Mie scattering code (Stackhouse 1991, private communication to SBDART authors) for spherical clouds droplets having a gamma size distribution and an effective radii, in the range 2 to 128 $\mu$m. (The effective radius is the ratio of the third and second moments of the droplet radius distribution). To allow analysis of radiative transfer through cirrus clouds the scattering parameters for spherical ice grains of a fixed size distribution with an effective radius of 106 $\mu$m have been also included.

### 4.2. Reflectance of the ground surface

The reflectance is a ratio of reflected to incident flux. The Bidirectional Reflectance Distribution Function (BRDF) specifies the behavior of surface scattering as a function of the illumination and view angles at a given wavelength:

$$\rho_\lambda(\theta_i, \theta_v, \phi_i, \phi_v) = \frac{dI_v(\theta_v \phi_v)}{\cos\theta_i dE(\theta_i \phi_i)}, \quad (33)$$

where $I_v$ is the emitted radiance, $dE = I_i d\omega$ is the irradiance (on the plane perpendicular to the light path) and $\theta_i, \phi_i, \theta_v, \phi_v$ are the polar angles which define the observation and incidence directions. The BRDF can be also expressed by the ratio of the radiance of the surface to the radiance of an ideal Lambertian surface $R_\lambda = \frac{\rho_\lambda}{1/\pi} = \pi\rho_\lambda$, which is more properly called Bidirectional Reflectance Factor (BRF). The albedo of a surface describes the ratio of the radiant energy scattered upward and away from the surface in all the directions to the down-welling irradiance incident upon the surface. The directional-hemispherical albedo in general depends on the zenith angle of incidence of the light and should not be confused with the bihemispherical albedo which is its average for any incident angle over the hemisphere. LPTRAN can obtain the spectral bidirectional reflectance and the albedo of the ground surface in the considered area from satellite data or from established models.

#### 4.2.1. BRDF from satellite data

NASA's Terra satellite and the MODerate Resolution Imaging Spectroradiometer (MODIS), provide global 1-km gridded and tiled multidate, multispectral products of the operational BRDF and albedo of the land surface with 16 days periodicity. The MODIS BRDF/Albedo Science Data Product is among the Level 3 1-km land products that are provided in an Integerized Sinusoidal Grid (ISG) projection with standard tiles representing 1200x1200 pixels on the Earth (Wolfe, Roy & Vermote 1998). The MOD43B MODIS BRDF/Albedo algorithm provides four standard products in HDF-EOS format for each of the 289 land tiles on the globe. The first two products supply spectral (MODIS channels 1–7) and broadband (0.3–0.7, 0.7–5.0, and 0.3–5.0 Am) BRDF model parameters so that the user can reconstruct the entire surface BRDF and compute the directional reflectance at any view or incident zenith angle desired.
The operational MODIS BRDF/Albedo algorithm makes use of a kernel-driven, linear BRDF model, i.e. the BRDF is expanded into a linear sum of functions (called kernels) of viewing and illumination geometry (Roujean, Leroy & Deschamps 1992):

$$R_\lambda(\theta_i, \theta_v, \phi) = f_{iso,\lambda} + f_{vol,\lambda} K_{vol}(\theta_i, \theta_v, \phi) + f_{geo,\lambda} K_{geo}(\theta_i, \theta_v, \phi), \quad (34)$$

where $\theta_i$ is the zenith distance of the incident light, $\theta_v$ is the zenith distance of the view direction, $\phi$ here is the azimuth angle of the view angle relative to the direction of incident light, $\lambda$ is the wavelength, $K_{vol}(\theta_i, \theta_v, \phi)$ and $K_{geo}(\theta_i, \theta_v, \phi)$ are the model kernels and $f_{iso,\lambda}$, $f_{vol,\lambda}$, $f_{geo,\lambda}$ are the spectrally dependent kernel weights. The kernel weights given by MODIS are those that best fit the available observational data. Several studies have identified the RossThickLiSparse- Reciprocal kernel combination as the model best suited for the operational MODIS BRDF/Albedo algorithm (Lucht, Schaaf & Strahler 2000; Privette, Eck & Deering 1997; Wanner, Li & Strahler 1995; Wanner et al. 1997).

The kernel for isotropic reflectance (Lambertian surfaces) is equal to one. The volumetric RossThick kernel $K_{vol}(\theta_i, \theta_v, \phi)$ accounts for volume scattering from an infinite discrete medium of randomly located facets and was derived from volume scattering radiative transfer models (Ross 1981) by Roujean et al. (1992):

$$K_{vol}(\theta_i, \theta_v, \phi) = \frac{(\pi/2 - \xi)\cos\xi + \sin\xi}{\cos\theta_i + \cos\theta_v} - \frac{\pi}{4}, \quad (35)$$

with,

$$\cos\xi = \cos\theta_i \cos\theta_v + \sin\theta_i \sin\theta_v \cos\phi. \quad (36)$$

The geometric LiSparseR kernel $K_{geo}(\theta, v, \phi, \lambda)$ accounts for the effects of shadows and the geometrical structure of protrusions from a Lambertian background surface and is derived from surface scattering and geometric shadow casting theory (Li & Strahler 1992). The surface is taken as covered by randomly placed projections (e.g. tree crowns) that are taken to be spheroidal in shape and centered randomly within a layer above the surface. For the LisparseR kernel, it is assumed that shaded crown and shaded background are black, and that lit crown and background are equally bright. The kernel was derived by Wanner et al. (1995):

$$K_{geo}(\theta_i, \theta_v, \phi) = O(\theta_i, \theta_v, \phi) - \sec\theta'_i - \sec\theta'_v + \frac{1}{2}(1 + \cos\xi')\sec\theta'_v \sec\theta'_i, \quad (37)$$

with,

$$\cos \xi' = \cos \theta'_i \cos \theta'_v + \sin \theta'_i \sin \theta'_v \cos \phi \quad (38)$$

$$O(\theta_i, \theta_v, \phi) = \frac{1}{\pi}(t - \sin t \cos t)(\sec \theta'_i + \sec \theta'_v) \quad (39)$$

$$\cos t = min\left[1, \frac{h}{b}\frac{\sqrt{D^2 + (\tan \theta'_i \tan \theta'_v \sin \phi)^2}}{\sec \theta'_i + \sec \theta'_v}\right] \quad (40)$$

$$t = \cos^{-1}(\cos t) \quad (41)$$

$$D = \sqrt{\left[\tan^2 \theta'_i + \tan^2 \theta'_v - 2\tan \theta'_i \tan \theta'_v \cos \phi\right]} \quad (42)$$

$$\theta'_v = \tan^{-1}\left(\frac{b}{r}\tan \theta_v\right) \quad (43)$$

$$\theta'_i = \tan^{-1}\left(\frac{b}{r}\tan \theta_i\right). \quad (44)$$

Here b is the vertical radius of the spheroid, r is the horizontal radius of the spheroid, and h is the height of the center of the spheroid. For MODIS processing is taken h/b=2 and b/r=1 (Strahler et al. 1999; see also Anon. 2003 for details).

*4.2.2. Model BRDF*

LPTRAN can also take the bidirectional reflectance from a choice of 8 accepted model functions, or a composition of them. The wavelength dependence of some of these models is given by the spectral reflectivity $\varpi_\lambda$ which is taken from SBDART. It uses six basic surface types: (i) sand (Staetter & Schroeder 1978); (ii) vegetation (Reeves, Anson & Landen 1975); (iii) ocean water (Viollier 1980); (iv) lake water (Kondratyev 1969); (v) clear water; (vi) snow (Wiscombe & Warren 1980). The spectral reflectivity of a large variety of surfaces can be approximated by combinations of these basic types. For example, the fractions of vegetation, water and sand can be adjusted to generate the spectral reflectivity representing new/old growth, or deciduous vs evergreen forest. Combining a small fraction of the spectral reflectivity of water with that of sand yields an overall spectral dependence close to wet soil. Other models require specific parameters for each wavelength. The BRDF models that we included in LPTRAN are the following:
(i) Bare soil. Hapke (1981) derived a model for dimensionless particles which provides a BRDF for soil:

$$R_\lambda(\theta_i, \theta_v, \phi) = \frac{\varpi_\lambda}{4}\frac{1}{\mu_v + \mu_i}$$
$$\times \left[(1 + B(\xi))P(g, \xi) + H(\mu_i, \varpi)H(\mu_v, \varpi) - 1\right], \quad (45)$$

where $\mu_i = \cos \theta_i$ and $\mu_v = \cos \theta_v$ and

$$B(\xi) = \frac{S(0)}{\varpi P(g, 0)}\frac{1}{\left[1 + (1/h)\tan(\xi/2)\right]} \quad (46)$$

$$P(g, \xi) = \frac{1 - g^2}{\left[1 + g^2 - 2g\cos(\pi - \xi)\right]^{1.5}} \quad (47)$$

$$H(x, \varpi) = \frac{1 + 2x}{1 + 2x\sqrt{(1 - \varpi)}} \quad (48)$$

$$\cos \xi = \cos \theta_i \cos \theta_v + \sin \theta_i \sin \theta_v \cos \phi. \quad (49)$$

$B(\xi)$ is a backscattering function that accounts for the hotspot effect, $S(0)$ defines the magnitude of the "hotspot", $P(g, \xi)$ is the Henyey and Greenstein phase function for the particle collection with asymmetry $g$, $H(x, \varpi)$ is a function to account for multiple scattering. The parameters that need to be supplied for a given land surface are: $\varpi$ the single scattering albedo, $g$ the asymmetry of the phase function, $S(0)$ is a parameter defining the height of the hotspot function at the hotspot, $h$ a parameter that controls the width of the hotspot function.
(ii) Vegetation: The Verstraete, Pinty, Dickinson (VPD) model (Verstraete, Pinty & Dickinson 1990; Pinty, Verstraete & Dickinson 1990) provides a BRDF for vegetation:

$$R_\lambda(\theta_i, \theta_v, \phi) = \frac{\varpi_\lambda}{4}\frac{k_i}{k_i\mu_v + k_v\mu_i}$$
$$\times \left[P_v(D)P(g, \xi) + H\left(\frac{\mu_i}{k_i}, \varpi_\lambda\right)H\left(\frac{\mu_v}{k_v}, \varpi_\lambda\right) - 1\right], \quad (50)$$

where $\mu_i = \cos \theta_i$, $\mu_v = \cos \theta_v$, $P(g, \xi)$ is the Henyey and Greenstein phase function given in eq. 16, $\cos \xi$ is given in eq. (18), and

$$k_x = \Psi_1 + \Psi_2\mu_x \quad (51)$$
$$\Psi_1 = 0.5 - 0.489\chi_l - 0.33\chi_l^2 \quad (52)$$
$$\Psi_2 = 1 - 2\Psi_1 \quad (53)$$
$$P_v(D) = \frac{1}{1 + V_p(D)} \quad (54)$$
$$V_p(D) = 4\left(1 - \frac{4}{3\pi}\right)\frac{D}{2r\Lambda}\frac{\mu_v}{k_v} \quad (55)$$
$$H(x, \varpi_\lambda) = \frac{1 + 2x}{1 + 2x\sqrt{(1 - \varpi_\lambda)}} \quad (56)$$
$$D = \sqrt{\left[\tan^2 \theta_i + \tan^2 \theta_v - 2\tan \theta_i \tan \theta_v \cos \phi\right]}. \quad (57)$$

The parameters that need to be supplied for a given land surface are: $\varpi_\lambda$ the single scattering albedo, $g$ the asymmetry of the phase function, $\chi_l$ the scatterer orientation parameter, $2r\Lambda$ the structural parameter (r is the sunfleck radius and $\Lambda$ is the scatterer area density).
(iii) Sea water. The bidirectional reflection of sea was obtained from SBDART v.2 (2002 release) sea water model, partly taken in turn from Tanre's 6S CODE (Vermote et al. 1995). It computes the reflectance of sea surface due to surface reflection, foam and subsurface particulates and Rayleigh scattering. It accounts for radiation directly reflected by the water surface as given by the Snell-Fresnel laws, salinity and chlorinity in the computation of the index of refraction and extinction coefficient of sea water, pigment concentration (e.g. chlorphyll+ pheophytin) in the computation of the spectral diffuse attenuation coefficient, wind speed and wind direction in respect to incident light which condition the model surface, computed numerically by many facets whose slopes are described by a Gaussian distribution which is considered anisotropic (depending upon wind direction).

(iv) Snow. The presence of snow can be recognized from daily snow cover data products available from the NASA/MODIS MOD10 snow and ice global mapping project. In particular, MOD10A1 and MOD10C1 are respectively snow cover daily global level 3 products with 500m ISIN grid and 0.05 deg CMG.(Hall et al. 2002; see also http://modis-atmos.gsfc.nasa.gov/). Reflectance of snow on soil varies consistently from fresh to senescent snow and BRDF shapes range from volumetric transmission with frontscattering to geometric crown with backscattering. As a rough empirical approximation, we adopted for fresh snow a combination of an isotropic kernel with a volumetric transmission kernel:

$$R_\lambda(\theta_i,\theta_v,\phi) = \varpi_\lambda \left(a + b\theta_i + c\theta_i^2\right)\left(1+\tau\frac{\sin\xi - \xi\cos\xi}{\cos\theta_i + \cos\theta_v}\right), \quad (58)$$

where $\xi$ was given in eq. (18), $\tau = 0.25$ is a transmission coefficient obtained with a rough fit to Painter & Dozier (2002) measurements of fresh snow, and $\left(a + b\theta_i + c\theta_i^2\right)$, with a=0.946, b=0.099, c=0.173, is a normalization function which makes the integral of $R/\pi$ equal to $\varpi_\lambda$ at any $\theta_i$ (which is not necessarily true). However a numerical BRDF for snow is also available from STREAMER package (Key, Schweige 1998; Key 2001).

We also included in the program (v) a Lambertian perfect diffuse reflector $R_\lambda(\theta_i,\theta_v,\phi) = \varpi_\lambda$, (vi) the RPV extended Minnaert model (Rahman et al. 1993), (vii) the Roujean et al. (1992) geometric + Ross-thick kernel model and (viii) a custom BRDF read from a user supplied file with the readbrdf subroutine from STREAMER package (Key et al.1998; Key 2001).

### 4.3. Upward light emission function

The normalized upward light emission function gives the relative intensity $\Im_{up}(\theta,\psi)$ emitted by a source at azimuth $\psi$ and zenith distance $\theta$, normalized so that its integral, the total upward flux, is unity (Cinzano et al. 2000, sec. 4.3). In computation of night sky brightness it should be multiplied by the upward flux per unit surface $e(x,y)$ or per unit land area $e_{i,j}$ or per source $e$, depending on the application. Any axisymmetric custom function $\Im_{up}(\theta)$ can be inserted in the code. However we adopted two series of functions of the kind $\Im_{up}(\theta) = \sum_{m=0}^{n} a_m \Im_{up,m}(\theta)$. Series are particularly useful because light pollution is additive. Computing separately the night sky brightness $b_m(\vartheta,\varphi)$, the scattered light $F_{i,j,k,p,q,m}$, and the radiation densities produced by each component $\Im_{up,m}(\theta)$, we can obtain these quantities for any upward function made by their linear combination:

$$b(\vartheta,\varphi) = \sum_{m=0}^{n} a_m b_m(\vartheta,\varphi) \quad (59)$$

$$F_{i,j,k,p,q} = \sum_{m=0}^{n} a_m F_{i,j,k,p,q,m}. \quad (60)$$

This is very useful in order to obtain reasonable computation times when computing a map of night sky brightness in territories where the upward emission function is not invariant or studying the effects of different upward emission functions, like, for example, those proposed by Luginbuhl et al. (2009). The two functions that we implemented in LPTRAN are (i) a 3-parameter upward emission function and (ii) a series of orthogonal Legendre polynomials.

*4.3.1. 3-parameter upward emission function*

A 3-parameter upward emission function was introduced as an extension of the Garstang (1986, eq.1) function by adding a mid-elevation emission centred at a zenith distance of 60. With only two shape parameters $u_2, u_3$ and one scale parameter $u_1$, this function can assume a number of shapes typically expected in upward emission functions. We first introduced it with the aim to recover the average upward light emission function in a territory from the measured brightness minimizing the number of free-parameters. The function is:

$$\Im_{up}(\theta) = u_1 \frac{2\cos\theta + u_2 0.5543\theta^4 + 1.778 q u_3 \cos(3\theta-\pi)}{2\pi(1+u_2+u_3)}. \quad (61)$$

with $q=0$ for $\theta < 30°$ and $\theta$ in radians. When $u_2 = 1.176$, $u_3 = 0$ it gives the Garstang function (Garstang 1986, eq.1) with the parameters adopted by Cinzano et al. (2000, eq. 15). When $u_2 = 0$, $u_3 = 0$ it gives a Lambertian emission. The shapes produced by other set of parameters will be shown in Cinzano, Falchi & Elvidge (in prep.), along with their effects on the sky brightness.

*4.3.2. Series of orthogonal Legendre polynomials*

Any function of $\theta$ can be expanded with a series of n orthogonal Legendre polynomials in function of $\cos\theta$ if n is sufficiently large that the residuum $O(n)$ is negligible:

$$\Im_{up}(\theta) = \sum_{m=0}^{n} p_m P_m(\cos\theta) + O(n), \quad (62)$$

where $P_m(\cos\theta)$ are the Legendre polynomials of degree m (Spherical Harmonics of the first kind with $\mu=0$) and $p_n$ are the coefficients which can be obtained for any function $I_{up}(\theta)$ by integrating, if necessary numerically, the equation:

$$p_n = \frac{2n+1}{2}\int_0^\pi \Im_{up}(\theta)P_n(\cos\theta)\sin\theta d\theta. \quad (63)$$

When only $p_1 \neq 0$ the result is a Lambertian emission function. This expansion allows the application of any function $\Im_{up}(\theta)$ passing to the program only the n coefficients $p_0,...,p_n$ of the series. It also allows the computation of the night sky brightness $b$, the array of the scattered light $F_{sca}$, and the radiation densities for any function $\Im_{up}(\theta)$ by summing those produced by each of the n Legendre polynomials times the corresponding coefficient as in eqs. (59) and (60). Care should be taken to check that the effects of the residuum of the expansion be negligible. The computation of the library requires a larger computational time with a Legendre series than with the 3-parameters emission function (n functions instead of three), but it allows the application of any expandable function.

## 5. RESULTS

Specific studies of the behaviour of light pollution in different atmospheric and surface conditions, or in specific sites, will be presented in forthcoming papers. In this paper we present a comparison between LPTRAN predictions and Garstang models and a sample of results which can be obtained with this method, including a map of the artificial sky brightness of regions of Chile where the main telescopes are located.

### 5.1. Comparison with Garstang models

A comparison between predictions of zenith artificial night sky brightness obtained with LPTRAN (solid line) and classic Garstang models (dashed line) is shown in fig. 1. Garstang models are computed for V-band, standard clear atmosphere, aerosol clarity K=1,3,5 (Garstang 1989). In order to mimic the Garstang model with K=1, LPTRAN prediction was obtained for 550 nm, US62 atmosphere, rural boundary layer aerosols with visibility 26 km corresponding to K=1, negligible ground reflection, no specific stratospheric aerosols and limiting the computation to the first two scatterings. We applied the 3-parameter upward light emission function with parameters reducing it to the function used by Cinzano et al. (2000, 2001a,b, 2004) and Garstang (1986, 1989) and normalizing it to unit upward light flux. We limited the computation to the first 120 km in this paper in order to reduce computational times but we plan to use 250-300 km in accurate applications. Fig. 1 shows that the LPTRAN prediction for a visibility of 26 km agrees closely with the corresponding Garstang model for K=1. As expected, differences arise when different atmospheric situations are considered. We added in the right panel of Fig. 1 a LPTRAN prediction for a visibility of 23 km, i.e. a larger aerosol content (dot-dashed line), for comparison.

We also plotted in the right panel of fig. 1 the generalized Walker Law $b = b_0 d^{-\alpha}$ (Walker 1977), where $b$ is the night sky radiance. Garstang (1986) and Joseph, Kaufman & Mekler (1991) showed that the exponent $\alpha$ depends on the aerosol content of the atmosphere, on the zenith and azimuth angles of the direction of observation and it become larger with the distance from the source. According to Joseph, Kaufman & Mekler (1991, Fig.6), the exponent $\alpha=2.5$ holds for distance from the source up to $\sim 30$ km, optical thickness $\tau \sim 0.25-0.3$ and brightness at zenith or at 45° toward the source. The Walker Law fits quite closely the LPTRAN and Garstang predictions for the adopted atmosphere in the range 8-120 km with $\alpha = 2.3(1+d/1000)$ (long-dashed line), where $d$ is the distance in km, and $b_0 \approx 2.1\cdot 10^{-3}$ sr$^{-1}$ km$^{-2}$.

We analyzed with LPDART the light pollution of the atmosphere in the situation described above. Fig. 2 shows the downward radiation density in the atmosphere as a function of the elevation and the distance from the source. This quantity shows the alteration of the natural light flux in each volume of atmosphere due to introduction of artificial light. In this figure and in the following two the abscissa shows the distance from the source in km and the ordinate shows the index number of the atmospheric volume. Up to 25 km it corresponds to the altitude of the top of the volume. From 26 to 32 it indicates respectively the volumes at altitudes between 25-30 km, 30-35 km, 35-40 km, 40-45 km, 45-50 km, 50-80 km, 80-100 km. The numbering sequence is inverted in respect to the original LPTRAN/LPDART output files. The units of this figure, and the next one, refer to a source with unit upward flux and are expressed in hundredths of decimal logarithm. The upper atmospheric layer was not plotted because the incoming downward radiation density is zero there, so that the logarithm is $-\infty$. Fig. 3 shows the light flux scattered downward by an unit volume of atmosphere as a function of the elevation and the distance of the source. It gives the "strength" of each volume of atmosphere as a secondary light source. Fig. 4 shows the fractional contribution of the light scattered by each unit volume of atmosphere to the artificial night sky brightness at zenith at sea level. Units are hundredths of per cent (e.g. 20.23% is given as 2023). Differently from the downward scattered flux in Fig. 3, this quantity accounts for the extinction along the path from the volume to the ground and it refers only to the light scattered along the nadir direction.

## 5.2. Light pollution in Veneto atmosphere

We computed with LPMAPALT/LPMAPDENS the light pollution in the atmosphere over the Veneto plane, Italy, assuming the atmospheric situation described above. This is only for test purposes and a more proper atmosphere will be adopted in specific studies. Input DMSP-OLS radiance data, GTOPO30 DEM data and calibration are the same as Cinzano, Falchi & Elvidge (2001b) and Cinzano & Elvidge (2004). We considered a rectangular area of 40'x95' in a latitude/longitude projection with the upper left corner at latitude N45°45' and longitude E10°54'. It is a considerably polluted area including, from West to East, the cities of Verona, Vicenza, Padova and Venezia.

Fig. 5 shows a map of the artificial night sky brightness at the sea level in $\mu$ cd/m$^2$, which can be compared with the predictions of the World Atlas of zenith night sky brightness at sea level (Cinzano, Falchi & Elvidge 2001a). Coordinates are in pixels 30" by 30" in size. Given that the night sky is defined as polluted where the artificial night sky brightness is larger than 10% of the natural night sky brightness (about $2.5 \cdot 10^2$ $\mu$ cd/m$^2$) the entire territory in the figure is polluted.

Fig. 6 shows a map of the horizontal illuminance at soil in $\mu$ lx produced by the artificial light scattered downward by the atmosphere. It looks slightly smoother than the night sky brightness of fig. 5. The horizontal illuminance produced by the natural sky luminosity at soil is of the order of $8 \cdot 10^2$ $\mu$ lx. If, in analogy with the night sky brightness, we define as polluted the soil where the artificial horizontal illuminance is larger than 10% of the natural one, then the soil of the entire territory in the figure is polluted. This figure gives only the illuminance due to artificial light scattered by the atmosphere and does not account for the direct illumination due to downward light wasted by lighting installations, which become the main component in the neighbourhood of the sources.

Given that 3D arrays cannot be printed in a figure, we present a section of the atmosphere above the Veneto plane along a line at constant latitude N46°26'. This line starts approximately from Verona, passes South of Vicenza and through the outskirts of Padova up to Venezia. Fig. 7 shows the downward radiation density in the atmosphere in Tb/km$^3$. The abscissa gives the position along the line in pixels and the ordinate gives the index number of the atmospheric volume in the original LPTRAN grid, which elevation in km is given in Table 1.

**Table1.** Altitude in km of the borders of each volume of atmosphere in LPTRAN/LPDART grid for this paper.

| 1 | 2 | 3 | 4 | 5 | 6 | 7 | 8 |
|---|---|---|---|---|---|---|---|
| 0-1 | 1-2 | 2-3 | 3-4 | 4-5 | 5-6 | 6-7 | 7-8 |
| 9 | 10 | 11 | 12 | 13 | 14 | 15 | 16 |
| 8-9 | 9-10 | 10-11 | 11-12 | 12-13 | 13-14 | 14-15 | 15-16 |
| 17 | 18 | 19 | 20 | 21 | 22 | 23 | 24 |
| 16-17 | 17-18 | 18-19 | 19-20 | 20-21 | 21-22 | 22-23 | 23-24 |
| 25 | 26 | 27 | 28 | 29 | 30 | 31 | 32 |
| 24-25 | 25-30 | 30-35 | 35-40 | 40-45 | 45-50 | 50-80 | 80-100 |

To quantify the alteration of the natural light flux in each volume of atmosphere due to the introduction of artificial light we can consider the atmosphere polluted when the artificial radiation density is greater than 10% of the natural one, as for the night sky brightness. The natural radiation density depends on the elevation but for clean atmosphere is roughly of the order of $2.6 \cdot 10^{-3}$ Tb/km$^3$, so that the atmosphere is polluted up to about 25 km. Fig. 8 shows approximately the altitude up to which the downward artificial radiation density of each atmospheric column over the considered territory is greater than the natural one. It should not be confused with the altitude up to which each atmospheric column is polluted, which is larger because it refers to 10% of the natural radiation density. Fig. 9 shows the fractional contribution to the artificial night sky brightness at zenith at sea level by each unit volume of atmosphere above the line of constant latitude N 46°26'. The integral from top of the atmosphere to the ground is unity.

## 5.3. Artificial sky brightness over northern Chile

We computed a map of zenith artificial sky brightness over Chile using the 2006 radiance calibrated data produced at NGDC/NOAA by Dr. Christopher Elvidge team. We used Lpmapalt applying the Garstang libraries with an atmospheric clarity of K=1 to maintain a continuity with the maps computed in the past. The used aerosol clarity of K=1 corresponds to a vertical extinction at sea level of Δm=0.33 mag in the V band, horizontal visibility of Δx=26 km, and optical depth τ=0.3. At 1000 m altitude the vertical extinction becomes Δm=0.21 mag and at 2000 m Δm=0.15 mag.

As seen in fig.1, the differences with LPTRAN libraries are absolutely not influent on the produced maps. For the computation of the sky brightness, we took into account the altitude of both the sources of pollution and observing sites, the screening of Earth curvature but not the screening due to mountains.

### 5.3.1 Calibration

We calibrated the maps using Earth based measurements of night sky brightness in the V band taken by one of us with the CCD based portable station (Falchi 2011). The calibration was checked also with data obtained with Sky Quality Meters (SQM-L) measurements taken in Chile by Pedro Sanhueza of OPCC for this purpose.

The uncertainty of calibration coefficients produces an uncertainty of about 20% in the predicted artificial brightness, in fact the standard deviation of the CCD measured values vs the predicted ones is 19%. Single sites may present far higher differences in the actual vs predicted brightness, due mainly to combinations of the uncertainty in the atmospheric transparency and conditions over the territory producing pollution in the site, the uncertainty and difference in the upward emission function of different cities, uncertainty and variations in the natural brightness subtracted from the measured data. This last factor is of great importance especially in sites with very low light pollution, where the signal (the artificial sky brightness) is much lower than the 'noise' (the natural sky brightness). Single data may differ from the prediction of the maps by as much as about 50%. Sites inside or very near great light pollution sources are subject to higher errors, mainly caused by the sky brightness dependence on the atmospheric clarity. For this reason sites inside cities were not used to calibrate the map. The calibration graph is shown in figure 10.

*5.3.2 Artificial sky brightness map*

The computed map is shown in figure 11. Levels represent the artificial sky brightness at zenith as the ratio to the natural sky brightness of an unpolluted night sky, supposed to be 22.0 mag arcsec$^{-2}$ at solar minimum, corresponding to 171 $\mu$cd m$^{-2}$. Colours correspond to ratios of <0.02 (black), 0.02-0.04 (dark grey), 0.04-0.08 (grey), 0.08-0.16 (dark blue), 0.16-0.32 (blue), 0.32-0.64 (dark green), 0.64-1.3 (green), 1.3-2.5 (yellow), 2.5-5 (orange), 5-10 (red), 10-20 (pink), 20-40 (magenta), >40 (white), so from one colour to the next there is approximately a doubling in the artificial brightness. As seen in the map, the observatories that are located in the vicinity of La Serena, while still not seriously polluted, are potentially more subject to the peril of a deleterious artificial sky brightening. Cerro Paranal and Cerro Armazones sites are, for now, in a safer situation and their skies are unpolluted at zenith. The map, computed using a standard upward emission function, does not take into account the positive effects that the complete enforcement of the rules against light pollution will have on all the observatory sites.

## 6. CONCLUSIONS

Light pollution propagates in the atmosphere, as other pollutants do, altering the involved medium and so it in not simply an alteration of the background for an observer of the night sky. We extended the seminal works of Garstang by providing a more general numerical solution for the radiative transfer problem applied to the propagation of light pollution in the atmosphere, which we called Extended Garstang Models (EGM). They retain the basic approach of Garstang models of computing the irradiance on each infinitesimal volume of atmosphere produced by the sources (including in this case secondary sources) and accounting for the extinction in the path. However EGM generalize the physical model with a more detailed computation of radiative transfer, Mie and Rayleigh scattering, line and continuous gas absorption, atmospheric and surface models.

We applied EGM to high-resolution DMSP-OLS satellite measurements of upward artificial light emissions and to GTOPO30 digital elevation data, which provides an up-to-date method to predict the artificial brightness distribution of the night sky at any site in the World at any visible wavelength for a broad range of atmospheric situations and the artificial radiation density in atmosphere across the territory. This constitutes an important extension of the methods presented by Cinzano & Elvidge (2004), Cinzano et al. (2000), Cinzano, Falchi & Elvidge (2001a,b) and will be used to compute a new world atlas of artificial sky brightness (Falchi, Cinzano, Elvidge, in prep.).

The software package LPTRAN to date is the state-of-the-art in computing artificial night sky brightness and in quantifying light pollution. Comparisons show that predictions of classic Garstang models and EGM fit closely when applied to clean sky, standard atmosphere and standard upward emission function. The programs of the LPTRAN package can accept as input atmospheric scattering properties or light pollution propagation functions obtained from any other source. However EGM account for (i) multiple scattering, (ii) wavelength from 250nm to infrared, (iii) Earth curvature and its screening effects, (iv) sites and sources elevation, (v) many kinds of atmosphere with possibility of custom setup (e.g. thermal inversion layers), (vi) mix of different boundary layer aerosols and tropospheric aerosols, with possibility of custom setup, (vii) up to 5 aerosol layers in upper atmosphere including fresh and aged volcanic dust and meteoric dust, (viii) variations of the scattering phase function with elevation, (ix) continuum and line gas absorption from many species, ozone included, (x) up to 5 cloud layers, (xi) wavelength dependent bidirectional reflectance of the ground surface from NASA/MODIS satellites, primary models or custom data (snow included), (xii) geographically variable upward emission function given as a three-parameter function or a Legendre polynomial series. A more general solution allows to also account for (xiii) mountain screening, (xiv) geographical gradients of atmospheric conditions, including localized clouds, (xv) geographic distribution of ground surfaces, but it suffers from computational requirements that are too heavy at this time. The price to pay for a more accurate behaviour is a slower computation of the libraries of light pollution propagation functions. Moreover, due to the constraint on computational times, the greater accuracy and detail introduced in the physical description is somehow counterbalanced by the need of larger approximations in the numerical computation, mainly related to the grid size. These limits will gradually disappear with time as faster computers become available.

A number of refinements can still be done in future years: (i) the replacement of spherical geometry with spherical refractive geometry can improve the predicted brightness at low elevation angles from distant sources; (ii) the computation of the natural night sky brightness with EGM rather than with the Garstang model (1989) will assure self-consistency to the computations of total sky brightness and limiting magnitude; (iii) the accurate determination of the angular distribution of nighttime light emissions from land areas obtained from OLS, future satellites or other methods will improve accuracy especially where laws against light pollution are enforced or where unusual lighting habits are applied; (iv) the availability of spectra of the light emission of each land area taken from satellites will allow an accurate prediction of the spectra of the polluted night sky; (v) the knowledge on a global scale of the change rates of light pollution from satellite data (Cinzano, Falchi, Elvidge in prep.) will make it possible to add time evolution to LPTRAN results; (vii) accurate measurements of night sky brightness together with accurate information on the related atmospheric situation will allow further testing and improvement of the modelling technique.

**ACKNOWLEDGMENTS**

We are indebted to Prof. Roy Garstang of JILA-University of Colorado for his friendly kindness in reading the first daft of this paper, for his helpful suggestions and for interesting discussions. We acknowledge Dr. Christopher Elvidge of NOAA/NGDC, Boulder, USA and the EROS Data Center, Sioux Falls, USA for kindly providing us the data used for test applications. We also thank Dr. Christopher Kyba for his contribution with helpful discussions and Dr. Paul Bogard for his help in English language.

Part of this research has been supported by the Italian Space Agency contract I/R/160/02 and by the Astronomy Department of the Universidad de Chile and the Office for the Protection of the Northern Skies of Chile – OPCC. The study of the light pollution in Veneto, Italy, belongs to a larger research project supported by the University of Padua CPDG023488. Asus TeK Italy kindly provided the computer used to compute the map of Chile.

**FIGURES**

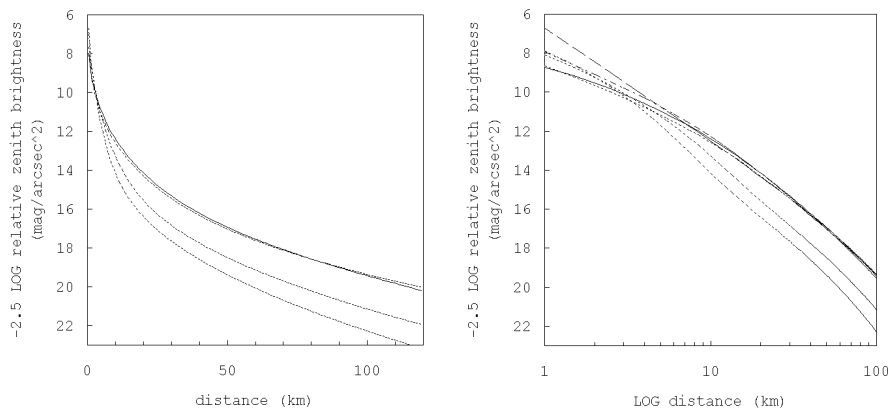

1. Artificial night sky brightness at zenith at sea level for the described atmosphere as a function of the distance from an unit source. Both panels show the LPTRAN prediction for a visibility of 26 km (solid line), corresponding to an atmospheric clarity K=1, and Garstang models for K=1,3,5 (dashed lines from top to bottom). Right panel also shows the LPTRAN prediction for a visibility of 23 km (dot-dashed line) and the generalized Walker Law with α=2.3(1+*d*/1000) (long-dashed line).

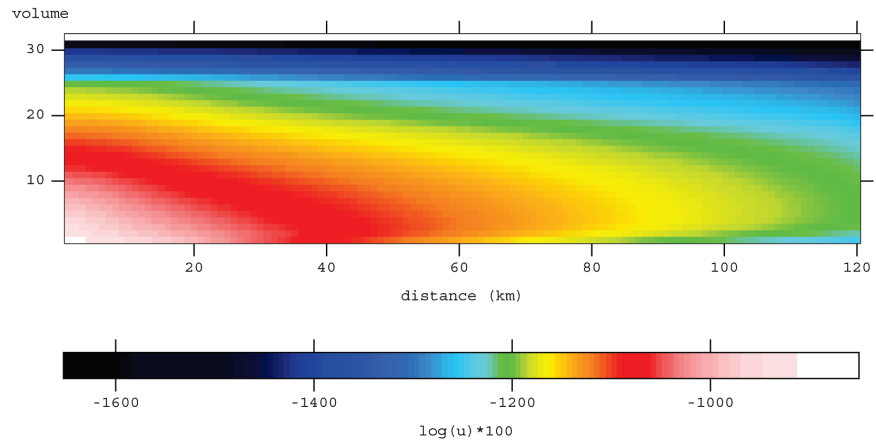

2. Downward radiation density in the atmosphere as a function of the volume index number and the distance from a unit source in the described case.

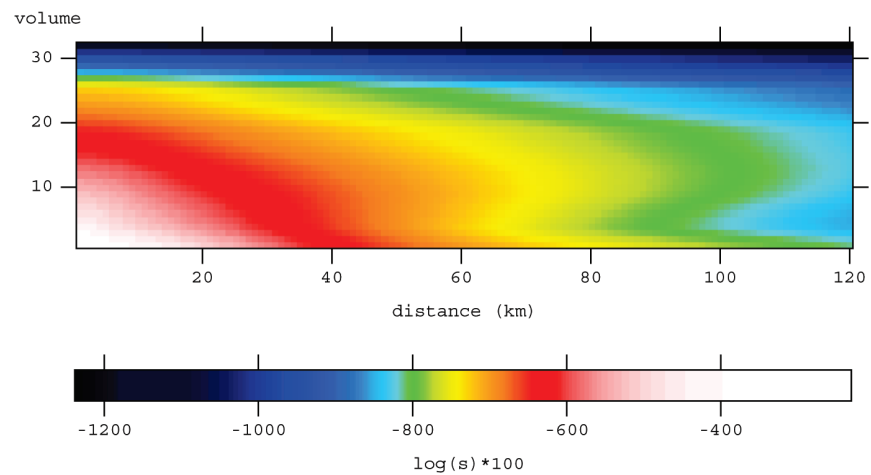

3. Light flux scattered downward by a unit volume of atmosphere as a function of the volume index number and the distance from a unit source for the described case.

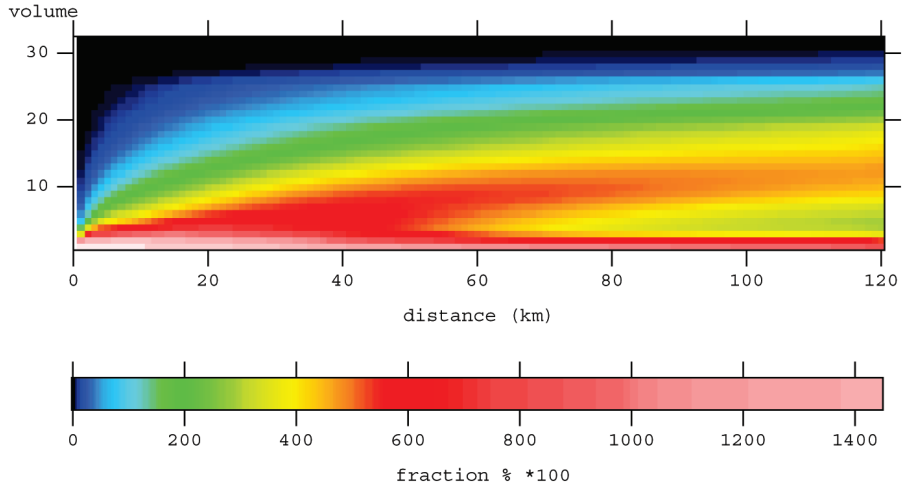

4. Fractionary contribution of the light scattered by each unit volume of atmosphere to the artificial night sky brightness at zenith at sea level for the described case.

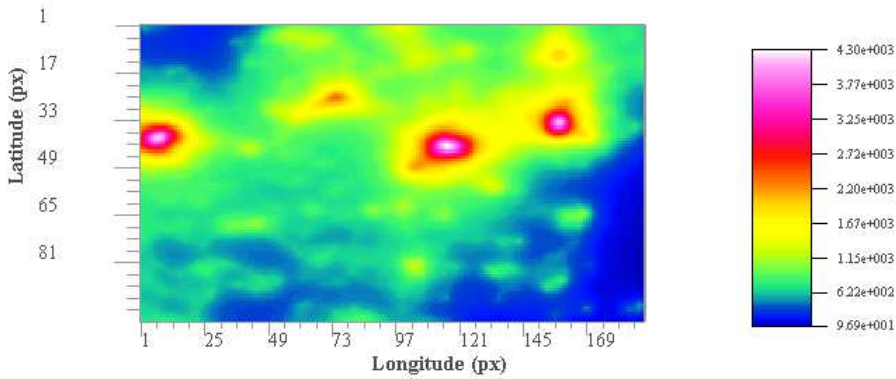

5. Artificial night sky brightness at the sea level in Veneto plane, Italy in $\mu$cd m$^{-2}$ for the atmosphere described.

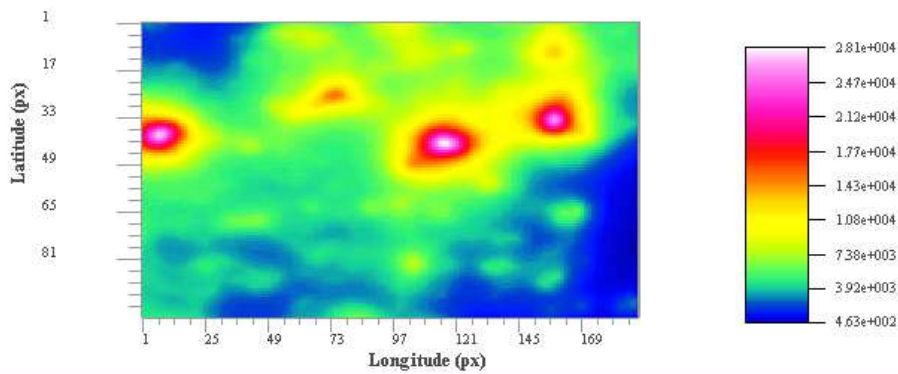

6. Horizontal illuminance at soil in Veneto plane, Italy in $\mu$lx, produced by the artificial light scattered downward by the atmosphere. The horizontal illuminance produced by the natural sky luminosity is of the order of $8\cdot 10^2$ $\mu$lx.

22    *P. Cinzano and F. Falchi*

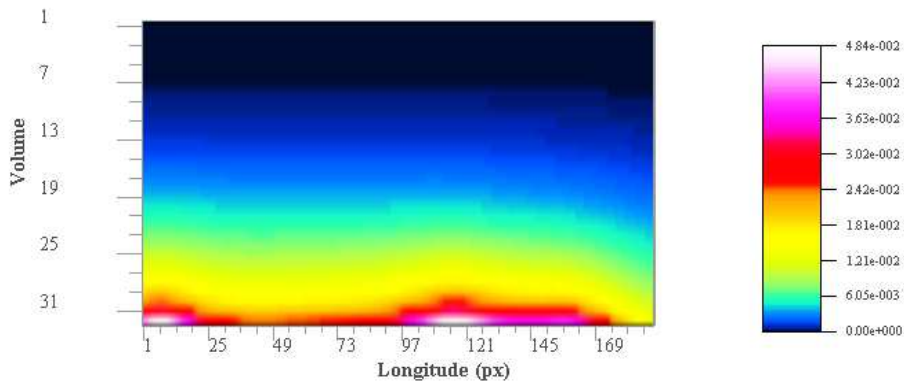

7. Downward radiation density in the atmosphere in Tb km$^{-3}$ above a line at constant latitude N46°26' in Veneto plane, Italy, as a function of the volume index number and the position. The natural radiation density is approximately of the order of $2.6 \cdot 10^{-3}$ Tb km$^{-3}$

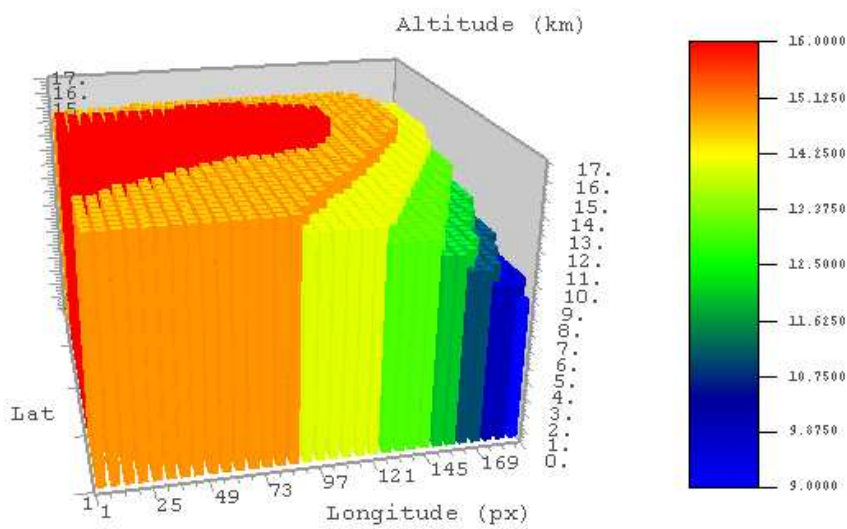

8. Altitude in km up to which the downward artificial radiation density of each atmospheric column over the considered territory is larger than the natural one. It should not be confused with the altitude up to which each atmospheric column is polluted, which is larger.

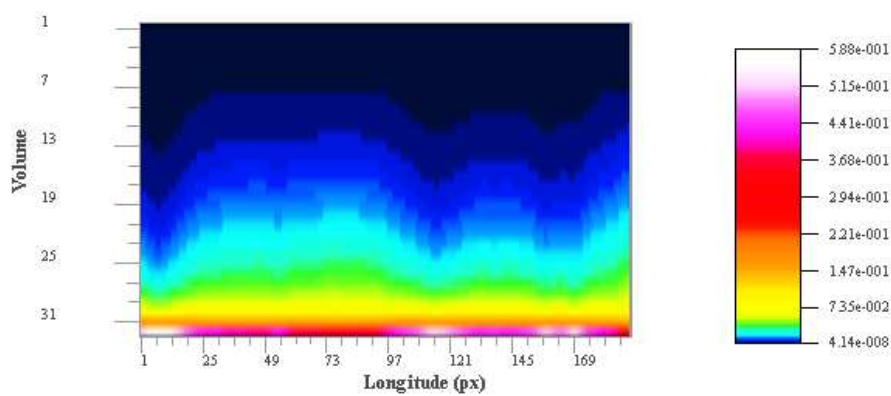

9. Fractional contribution to the artificial night sky brightness at zenith at sea level by each unit volume of atmosphere above the considered line of constant latitude N46°26'. The integral from top of the atmosphere to the ground is unity.

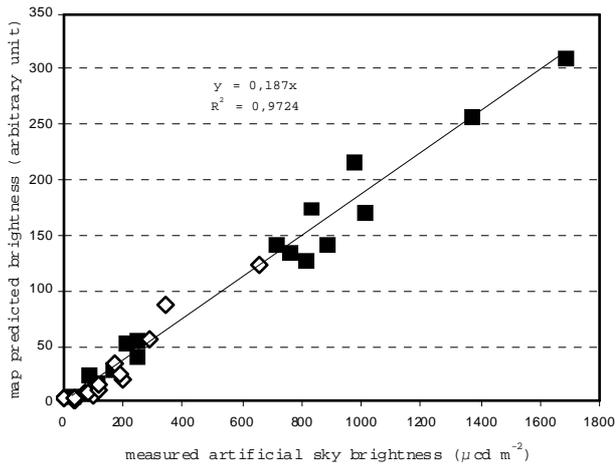

10. Calibration graph for the map of Figure 12. Black squares are the V brightness CCD measures used for the interpolation line and the subsequent calibration. The diamonds are the measurements taken with SQM-L in Chile to check the calibration.

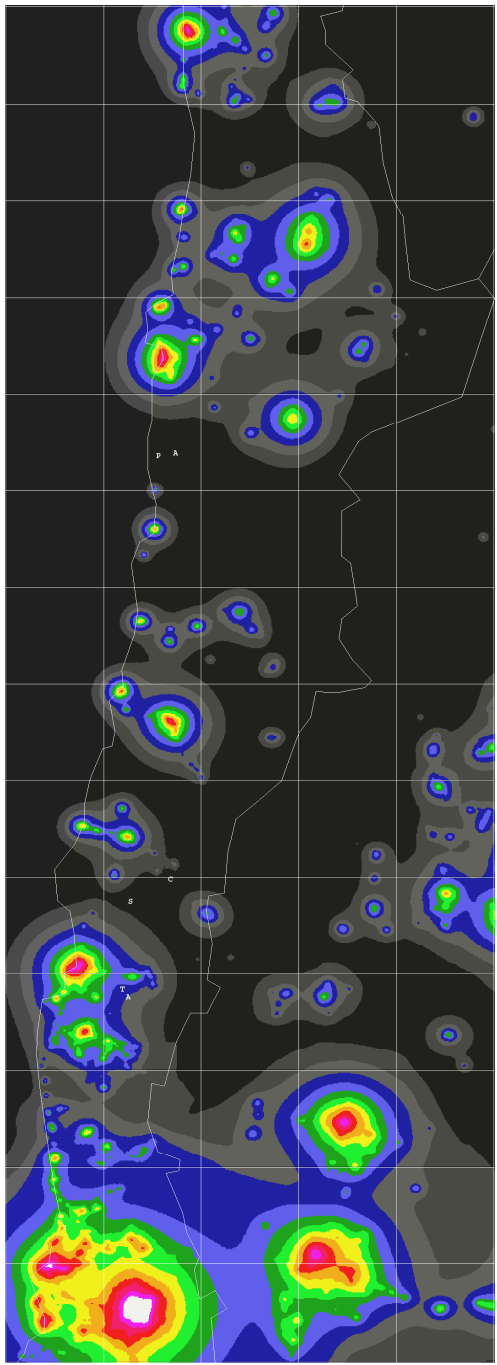

11. Artificial night sky brightness in north and central Chile, taking into account for elevation. Colours indicate, in twofold increases, the artificial sky brightness expressed as a ratio with the natural sky brightness at solar minimum (assumed to be 22.0 mag·arcsec$^{-2}$ in the V band, corresponding to 171 $\mu$cd m$^{-2}$): <0.02 (black), 0.02-0.04 (dark grey), 0.04-0.08 (grey), 0.08-0.16 (dark blue), 0.16-0.32 (light blue), 0.32-.64 (dark green), 0.64-1.3 (green), 1.3-2.5 (yellow), 2.5-5 (orange), 5-10 (red), 10-20 (pink), 20-40 (magenta), >40 (white). The grid is 1°x1° in latitude-longitude projection with the lower left corner at 34° S, 72° W. White letters indicate the approximate positions of the main optical observatories: T= Cerro Tololo, P= Cerro Pachon, S= La Silla, C= Las Campanas, *P (inclined)* = Cerro Paranal, A= Cerro Armazones.

24  *P. Cinzano and F. Falchi*